# Silver nanowires with optimized silica coating as versatile plasmonic resonators

Martin Rothe,[1] Yuhang Zhao,[2] Günter Kewes,[1] Zdravko Kochovski,[2] Wilfried Sigle,[3] Peter A. van Aken,[3] Christoph Koch,[4] Matthias Ballauff,[2,5] Yan Lu,[2,6] and Oliver Benson[1]

[1] Humboldt Universität zu Berlin & IRIS Adlershof, Nanooptics, Newtonstraße 15, 12489 Berlin, Germany
[2] Helmholtz Zentrum Berlin für Materialien und Energie, Institute of Soft Matter and Functional Materials, Hahn-Meitner-Platz 1, 14109 Berlin, Germany
[3] Stuttgart Center for Electron Microscopy, Max Planck Institute for Solid State Research, Heisenbergstr. 1, 70569 Stuttgart, Germany
[4] Humboldt Universität zu Berlin & IRIS Adlershof, Structure Research and Electron Microscopy, Newtonstraße 15, 12489 Berlin, Germany
[5] Humboldt Universität zu Berlin, Department of Physics, 12489 Berlin, Germany
[6] Institute of Chemistry, University of Potsdam, 14467 Potsdam, Germany

## Abstract

Metal nanoparticles are the most frequently used nanostructures in plasmonics. However, besides nanoparticles, metal nanowires feature several advantages for applications. Their elongation offers a larger interaction volume, their resonances can reach higher quality factors, and their mode structure provides better coupling into integrated hybrid dielectric-plasmonic circuits. It is crucial though, to control the distance of the wire to a supporting substrate, to another metal layer or to active materials with sub-nanometer precision. A dielectric coating can be utilized for distance control, but it must not degrade the plasmonic properties. In this paper, we introduce a controlled synthesis and coating approach for silver nanowires to fulfill these demands. We synthesize and characterize silver nanowires of around 70 nm in diameter. These nanowires are coated with nm-sized silica shells using a modified Stöber method to achieve a homogeneous and smooth surface quality. We use transmission electron microscopy, dark-field microscopy and electron-energy loss spectroscopy to study morphology and plasmonic resonances of individual nanowires and quantify the influence of the silica coating. Thorough numerical simulations support the experimental findings showing that the coating does not deteriorate the plasmonic properties and thus introduce silver nanowires as usable building blocks for integrated hybrid plasmonic systems.

# Introduction

In plasmonics, field enhancement effects near metallic nanostructures are utilized to strengthen the light-matter interaction and to miniaturize optical or optoelectronic functionality. Most prominently, plasmonic resonances of metal nanoparticles (NPs) have been used for a plethora of applications and are well understood both experimentally and theoretically[1-11]. However NPs feature some fundamental limitations, e.g., concerning their minimal damping rates or quality ($Q$) factors[10]. Further, frequently studied plasmonic systems are composites of emitters and nanostructures, e.g., for strong coupling[8,12] or lasing[13-15]. In both scenarios, the envisioned functionalities can be limited by the small physical size of the nanoresonator's near-field zone, i.e., preventing to host a sufficient number of emitters. Nanowires hold more potential for designing customized nanostructures with both, relatively high Q-factors and yet still small mode volumes, respectively[16,17]. Thus, waveguiding nanostructures, like nanowires (NWs) might be preferred over NPs in many situations. Adjusting cross section and length of NWs allows for intuitive tuning and optimization towards specific needs such as higher Q-factors that go beyond the limits of NPs[18]. Furthermore, plasmonic waveguide structures might be much better suited for integration into dielectric on-chip waveguide structures[19-21].

Silver nanowires of finite length are especially interesting since they represent ideal type Fabry-Pérot resonators[22,23]. Also silver is well-known for its superior plasmonic properties in the visible spectral range[24]. With respect to the aforementioned composite systems of emitters and nanostructures, it is crucial to control distances on a nm-scale. Besides potential enhancement, metallic nanostructures could also negatively affect nearby emitters due to additional non-radiative channels. The same nm-control is needed to define the distance to nearby materials like additional metallic nanostructures or flat surfaces, e.g., in a wire-on-metal geometry. The surrounding materials shape the mode pattern of plasmonic resonances or guided modes[17], affecting the field localization as well as damping rates. Also for coupling to integrated dielectric structures, e.g., via directional coupling[25] or special plasmon-photon transducers[26] a high accuracy is mandatory. A promising method to reach this goal of precise distance control is a well-defined coating around the metallic nanostructures. Furthermore, for the realization of emitter–resonator composites, a porous dielectric coating like amorphous silica can act as a host for the emitters.

Amorphous silica has been proven an excellent candidate for coating nanostructures, as it is optically transparent, chemically inert and photo-chemically stable[27]. It is also perfectly suited for both emitter-hosting and distance control. While coating with silica shells has been frequently applied to NPs made from gold [27-30] during the past decade, less work has focused on coating of silver nanostructures like nanowires with amorphous silica shells[31,32]. Silver, however, would provide superior plasmonic properties at shorter wavelengths typical for active nanostructures or in single emitter (quantum) plasmonics[33]. Further, in the reported approaches, the silver is likely to be affected by the used ammonia solution.

In the following, we introduce a novel approach based on a modified Stöber method that improves on these shortcomings for higher quality of the coated silver wires as well of the silica shells. We report on the chemical synthesis and characterization of silver nanowires with a controlled silica-coating, which is well defined in terms of thickness, homogeneity and especially in surface roughness as compared to reports in literature[32,34,35]. The geometry of the silver nanowires is characterized by transmission electron microscopy while UV-VIS, dark-field scattering spectroscopy (DF) and electron energy-loss spectroscopy (EELS) are used to study the influence of the silica coating on the surface plasmon resonances. We measure the effective refractive index of the guided plasmon mode (or the momentum of the surface plasmon polariton (SPP)) and compare it to bare silver nanowires and nanowires with an 8 nm thick silica coating. Further, we employ a Fourier-transformation-based algorithm that allows gaining information about plasmon propagation and damping[36-38]. Finally, we support our findings with numerical Maxwell simulations observing perfect agreement.

## Results and discussion

**Synthesis and structural analysis**

Silver nanowires were synthesized via a two-injection step in a polyol process.[39-41] Figure 1a shows a typical bright-field TEM image of the as-prepared bare silver nanowires. As can be seen from the image, homogeneous silver nanowires with average length of 5 µm and diameter of 70 nm were fabricated. The distribution histogram of the length of the silver nanowires is shown in Fig. 1d. The inset of Fig. 1b presents an electron diffraction pattern obtained by reducing the size of the electron beam to illuminate only one nanowire, indicating that a twinned crystal structure forms the silver nanowire. The high-resolution TEM images in Fig. 1c reveal that each portion of this twinned nanowire is single-crystalline, and features well-resolved interference fringe spacing.[42,43]

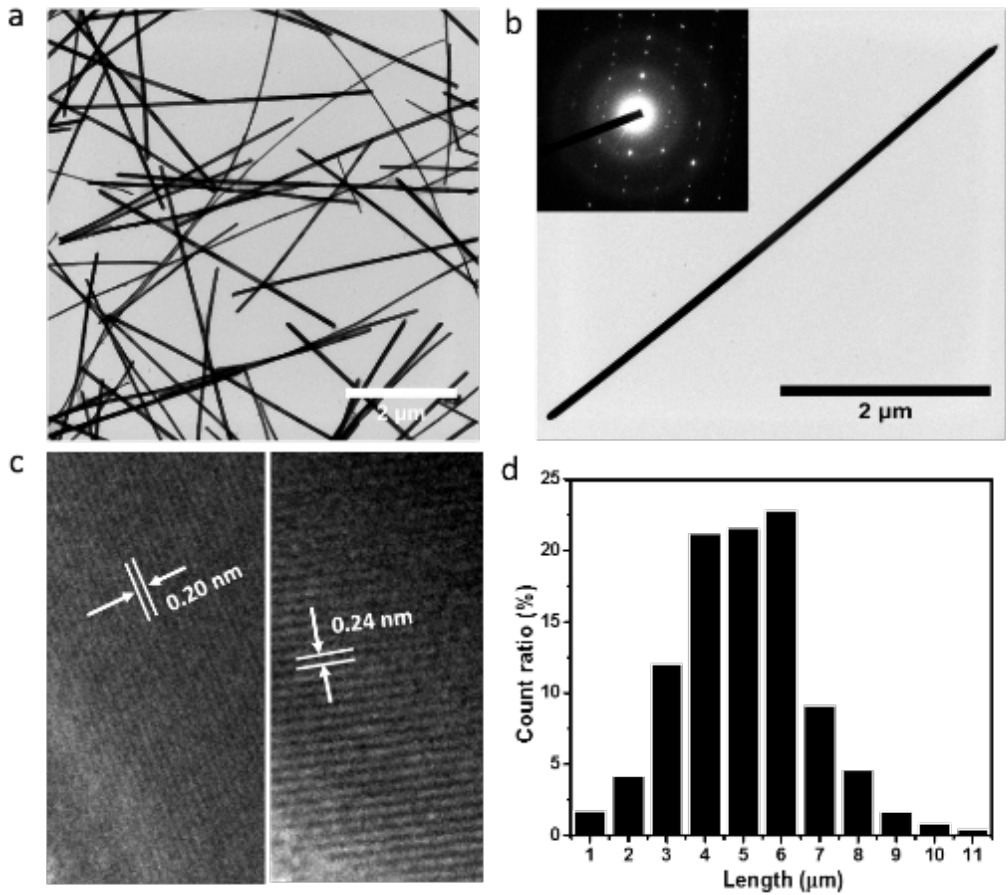

Figure 1. (a) TEM image of pure silver nanowires. (b) TEM image of a single pure silver nanowire, the inset is the corresponding selected-area electron diffraction pattern. (c) High-resolution TEM images taken from each edge of this silver nanowire, indicating the single crystallinity of each side. Lattice spacing of 0.2 nm and 0.24 nm correspond to (200) and (111) planes, respectively. (d) Length-distribution histogram of pure silver nanowires counted from TEM micrographs for a particular area.

In a next step, we address the coating procedure. So far, coating of the silver nanowires with an amorphous silica shell have been prepared via the Stöber method.[31] In a typical procedure, the encapsulation by a silica shell involves the hydrolysis of tetraethyl orthosilicate (TEOS) and subsequent formation of silica surrounding the silver nanowires. Normally, the hydrolysis of TEOS is performed under an alkaline environment. In the literature, ammonia solution is widely being used to serve as the alkaline condition[44,45]. In our study, we have first coated silver nanowires with an approximately 8 nm thick silica shell assisted by an ammonia solution, as shown in Fig. 2a and Fig. 2b. By increasing the concentration of TEOS from 0.26 vol% to 0.52 vol%, the silica shell thickness increases from around 8 nm to 24 nm. However, although a silica coating was formed by the assistance of the ammonia solution, one can clearly see from the TEM images that the ends of the silver nanowires have been etched by the ammonia solution. Further, also the surface of the silica shells appears wrinkled and uneven. We conclude that ammonia attacks the silver nanowires through the following reaction:[45]

$$4Ag + O_2 + 8\,NH_3 \cdot H_2O \rightarrow 4Ag(NH_3)_2^+ + 4OH^- + 6H_2O.$$

These structural defects might prejudice the optical potential of the well-designed core-shell silver nanowires-based nanostructure. To avoid this phenomenon, we replaced the ammonia solution by sodium hydroxide to provide an alkaline environment for the formation of homogeneous silica shell structures. As can be seen in Fig. 2d and e, silver nanowires with around 8 nm-thick silica shells could successfully be prepared.

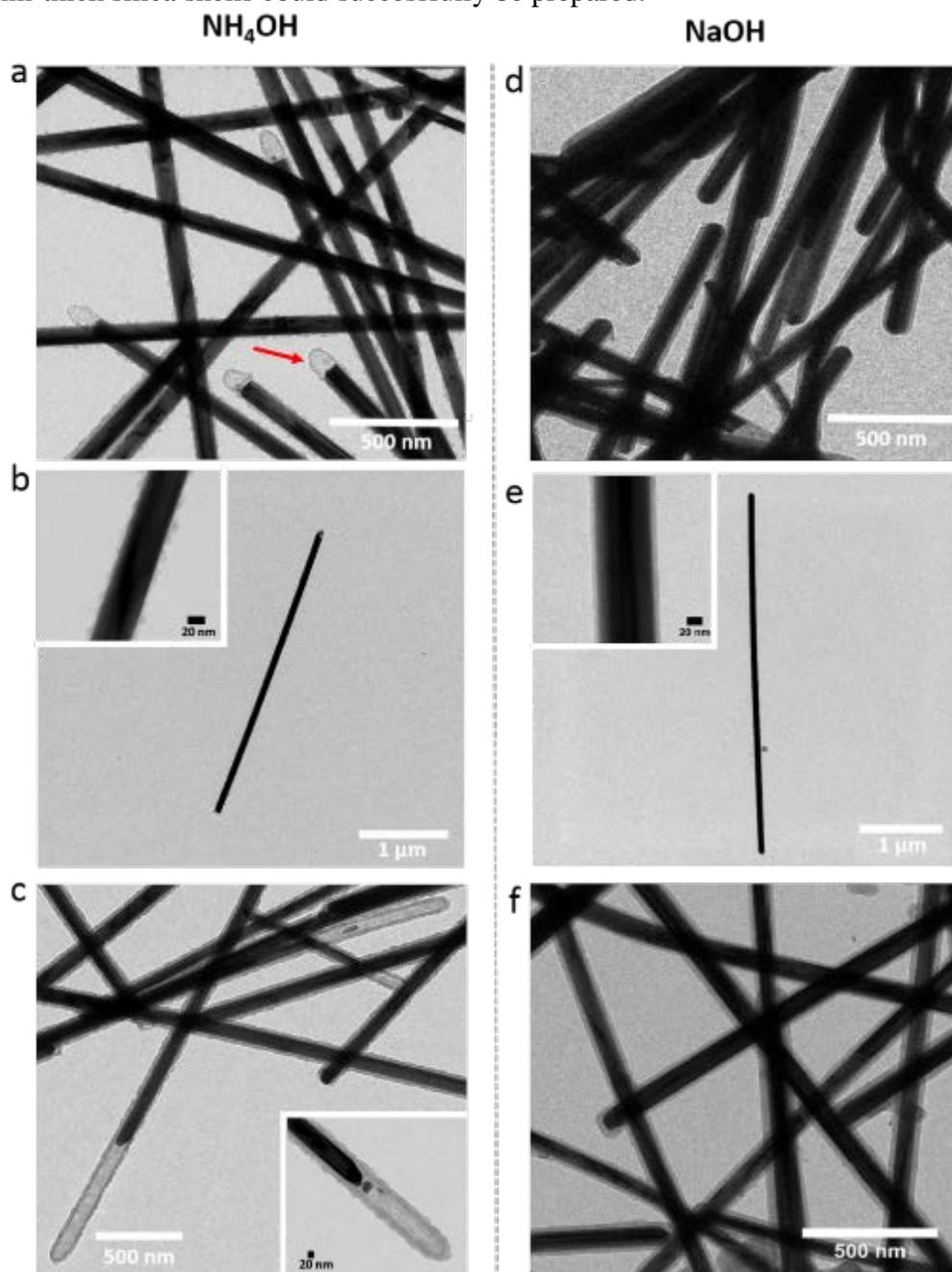

Figure 2. Left column: TEM images of silver nanowires coated with an 8 nm thick silica shell synthesized in ammonia solution: A-SiO$_2$ 8nm@AgNWs (a) overview image, (b) a single A-SiO$_2$@AgNWs, the inset shows a magnified insight into the inhomogeneous silica coating

structure. (c) TEM overview image of A SiO$_2$24nm@AgNWs synthesized in ammonia solution, the inset shows a zoom-in to the etched end of the silver nanowires in silica coating. Right column: TEM images of silver nanowires coated with an 8 nm silica shell synthesized in sodium hydroxide: SiO$_2$8nm@AgNWs (d) overview image, (e) a single composite of SiO$_2$8nm@AgNWs, the inset shows a magnified insight into the smooth silica coating structure. (f) TEM overview images of SiO$_2$24nm@AgNWs synthesized in sodium hydroxide solution.

As in the case of bare nanowires, single, coated wires are easily identified. Via the improved method, the as-prepared silica surface is very smooth and the silver core of the nanowires maintains its original structure. By increasing the TEOS concentration from 0.26 vol% to 0.52 vol%, silica shells with a thickness of 24 nm were achieved uniformly covering the entire surface of the silver nanowire without etching the silver nanowire's core structure (see Fig. 2f). This indicates that an excellent control of the thickness of the silica shell can be achieved by changing the concentration of TEOS. These results highlight the potential of the improved silica coating method to achieve silver nanowires with homogeneous and thickness-controllable silica shells.

We now focus on studying the optical properties by means of UV-VIS ensemble measurements of the as-grown nanowires and of the coated nanowires using ammonium and sodium hydroxide. Fig. 3 shows the UV-VIS spectra of silver nanowires dispersed in aqueous solution with a standard ultraviolet-visible absorption of pure silver nanowires showing features of longitudinal and transversal SPPs as well as bulk plasmons.[46] The peak at 350 nm is attributed to a bulk response[23], while the other one at 382 nm is ascribed to the transversal plasmon resonance (excitation with light polarized along the short axis of the wires), respectively.[47] We conclude that the broad shoulder starting at 382 nm and reaching towards around 650 nm represents the averaged spectral response of the ensemble of nanowires with different length. For the coated AgNWs we find broadened spectral features that are most prominent for the thickest coating of around 25 nm. For the thinner coating of around 8 nm we find that the ammonium treated AgNWs show a modified bulk plasmon peak, while the sodium hydroxide treatment leaves this peak rather unaffected. Thus, we take these ensemble measurements as first indication that the sodium hydroxide indeed represents a gentler treatment than the treatment with ammonium.

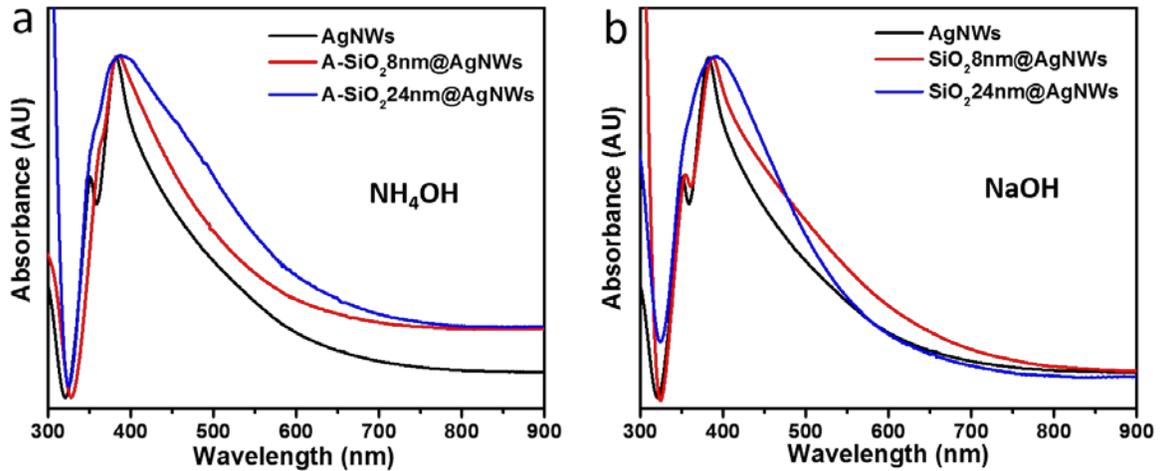

Figure 3. UV-VIS absorption spectra of pure silver nanowires and silica coated silver nanowires synthesized in different alkaline environment: (a) ammonia solution, (b) sodium hydroxide. The peaks at 350 nm and 382 nm are ascribed to the bulk and the transversal plasmon, respectively.

**Study of individual AgNWs**

To study the quality of the sodium hydroxide treated AgNWs in more detail we now turn to our studies on individual AgNWs. A clear indication of a good quality in plasmon resonators is the appearance of standing wave resonances. For this reason we look for and analyze such Fabry-Pérot-type resonances in our silver wires. In order to observe the influence of the silica coating on the plasmonic mode of individual silver nanowires, silver nanowires with 8 nm and 24 nm silica coating were prepared on cover glass by spin coating, respectively. The resulting randomly distributed nanowires could then be studied one-by-one in our home-build microscope setup (Fig. 4).

Bright-field imaging (Fig. 5a inset) allows for detection of single nanowires. Subsequently we use dark-field illumination with linear polarized light from a Xenon arc lamp to analyze the scattering properties of the nanowires. We focus on those nanowires, which lie aligned parallel to the direction of propagation of the incident light, as these show strong scattering only from their front and rear facet and thus less background (Fig. 5a). We spatially separate light from each end facet by the confocal principle and guide it to the spectrometer. The obtained scattering spectra were background subtracted and normalized to the measured and background-subtracted spectrum of the white light source.

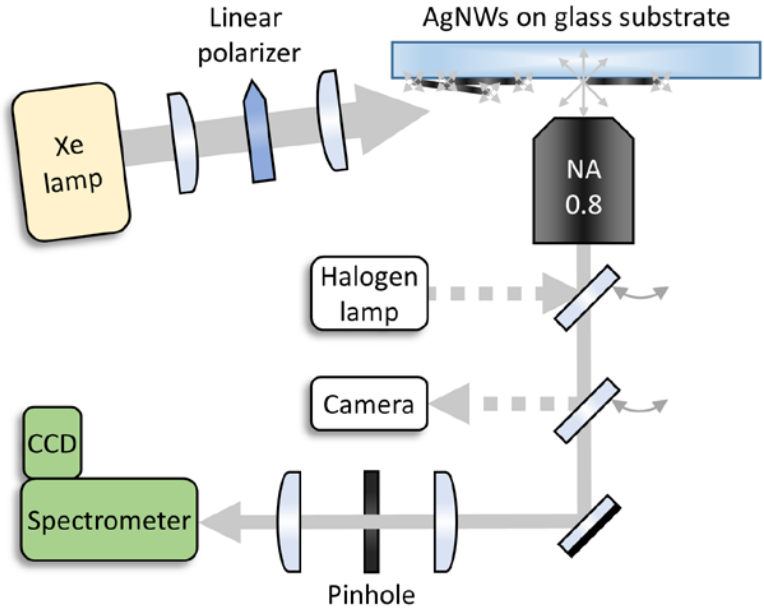

Figure 4. Experimental setup: Optical measurements on single nanowires with a home-build microscope. A halogen lamp is used for bright-field illumination. A Xenon arc lamp provides white light, which is subsequently linearly polarized and sent from the side to create a dark-field configuration. Scattered light is filtered spatially by focusing on a pinhole and detection by a spectrometer.

Typical normalized scattering spectra from a single silver nanowire are shown in Fig. 5b-e. Selecting the polarization of the incident light allows addressing distinct plasmonic modes. Furthermore, one has to distinguish between the front and the rear facet scattering, as additional interference effects might be observed that are not necessarily stemming from the Fabry-Pérot interferometer formed by the individual AgNWs. For p-polarized illumination, the spectrum of the front facet shows strong oscillations (Fig. 5d), that is a clear signature of coupling between the white light and the longitudinal plasmon modes. We find that the spectra can be modeled as the reflection of a Fabry-Pérot resonator[38,48]. According to the relatively strong damping in such a plasmonic resonator the expected spectral features like peaks and dips broaden and follow a sinusoidal behavior. When the polarization is changed to s-polarization (Fig. 5b and Fig. 5c) the oscillation vanishes towards a broad peak in the blue range of the spectrum. In this case, no coupling to longitudinal modes is expected due to the symmetry of these modes. The broad peak stems from scattering by the short axis of the wire, i.e. it is a signature of the transversal mode. This feature is also present in the spectra recorded with p-polarization as the used polarization filter has limited performance for the blue side of the spectrum.

At the rear facet we find again an oscillating spectrum when using p-polarized illumination (Fig. 5e), however with an additional modulation. This is due to a superposition of two contributions caused by the simultaneous illumination of both facets. One contribution is the direct scattering from the rear facet, the other one stems from light that was transmitted through the wire and then eventually scattered into photonic modes. Both, reflection and transmission spectra of a Fabry-Pérot resonator carry the same information about the resonator. Since the above mentioned superposition does not occur in the front facet spectra, these spectra are used for further analysis of the resonator. Our analysis yields the real part of the effective refractive index Re($n_{\text{eff}}$) of the guided mode and the plasmon round trip loss. The latter contains propagation losses α and losses due to imperfect end facet reflectivities *R*.

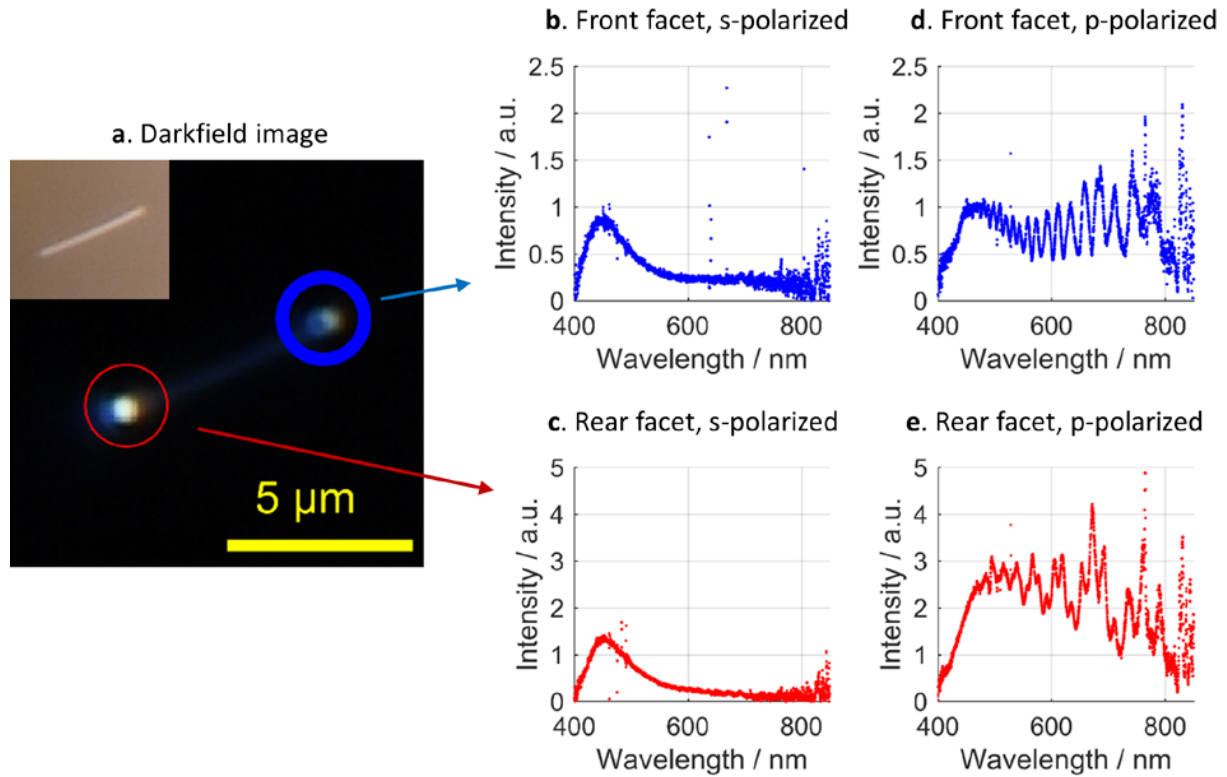

Figure 5. A single silver nanowire detected by dark-field (a) and bright-field (inset) imaging. Both, the front facet (b) and the rear facet (c) show a broad resonance in the scattering spectrum under s-polarized illumination. Under p-polarized illumination, longitudinal plasmon modes are excited. Thus, the scattered light reveals Fabry-Pérot resonances most clearly visible for scattering at the front facet (d). The rear facet (e) shows similar features, but overlaid with interference with the incident light.

While dark-field spectroscopy in principle works nicely for nanoparticles, it may not unambiguously exclude any source of interference that might only mimic Fabry-Pérot resonances for larger structures[49,50]. Thus, we further performed complementary EELS measurements of AgNW specimen, as EELS directly provides maps of the projected local optical density of states of standing wave patterns of the resonating field along the AgNW resonator. Figure 6 summarizes our findings where we recorded a set of EELS spectra for varying positions along coated and as-grown nanowires.

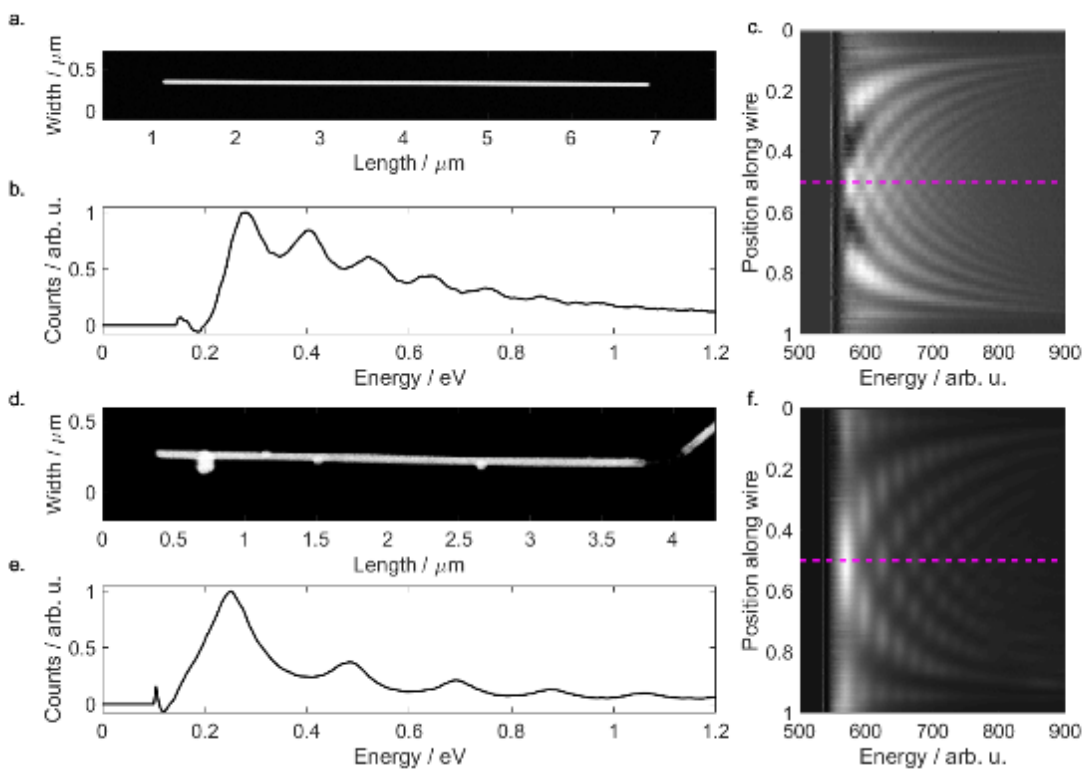

Figure 6. Summary of EELS measurements of an as-grown nanowire (a-c) and a coated nanowire with 8 nm silica shell(d-f). (a,d) TEM images of nanowires, (b,e) EELS spectrum recorded next to the nanowire's center as indicated by the dashed line in c) and f), respectively. (c,f) Spectral EELS maps recorded along the nanowires: horizontal lines correspond to spectra at fixed positions, while vertical lines correspond to standing wave pattern for fixed loss energies.

**Quantitative analysis of plasmonic losses**

We now move on with a quantitative analysis of our data. The most important parameter to judge the quality of our plasmonic nanostructures is the loss rate or round trip loss (RTL). We experimentally derive this number by analyzing the spectra utilizing a fast Fourier transformation (FFT). Figure 7 presents bright-field and dark-field images together with scattering spectra and FFT spectra, for example of an as-grown AgNW (Fig. 7a), a AgNW with 8 nm silica coating (Fig. 7d) and a AgNW with 24 nm silica coating (Fig. 7g). We point out here that the preparation of the nanowires with 24 nm shells was difficult due to strong clustering of this type of nanowire in dispersion before the spin coating preparation. Stronger and longer sonification dissolved the clusters, but also led to shorter AgNW due to NW fracture.

We selected a wavelength range of roughly 100 nm around the center wavelength of 650 nm (white area in the second column of Fig. 7) for further analysis, i.e., the spectral range for which we found the best signal to noise ratio. The selected range was converted to the frequency space and a FFT was performed to analyze the resonator properties round trip time and damping.[36-38] The boundaries of the spectral window were selected to coincide with a local minimum or maximum to improve the quality of the FFT. Furthermore, only spectra with at least two oscillations where considered.

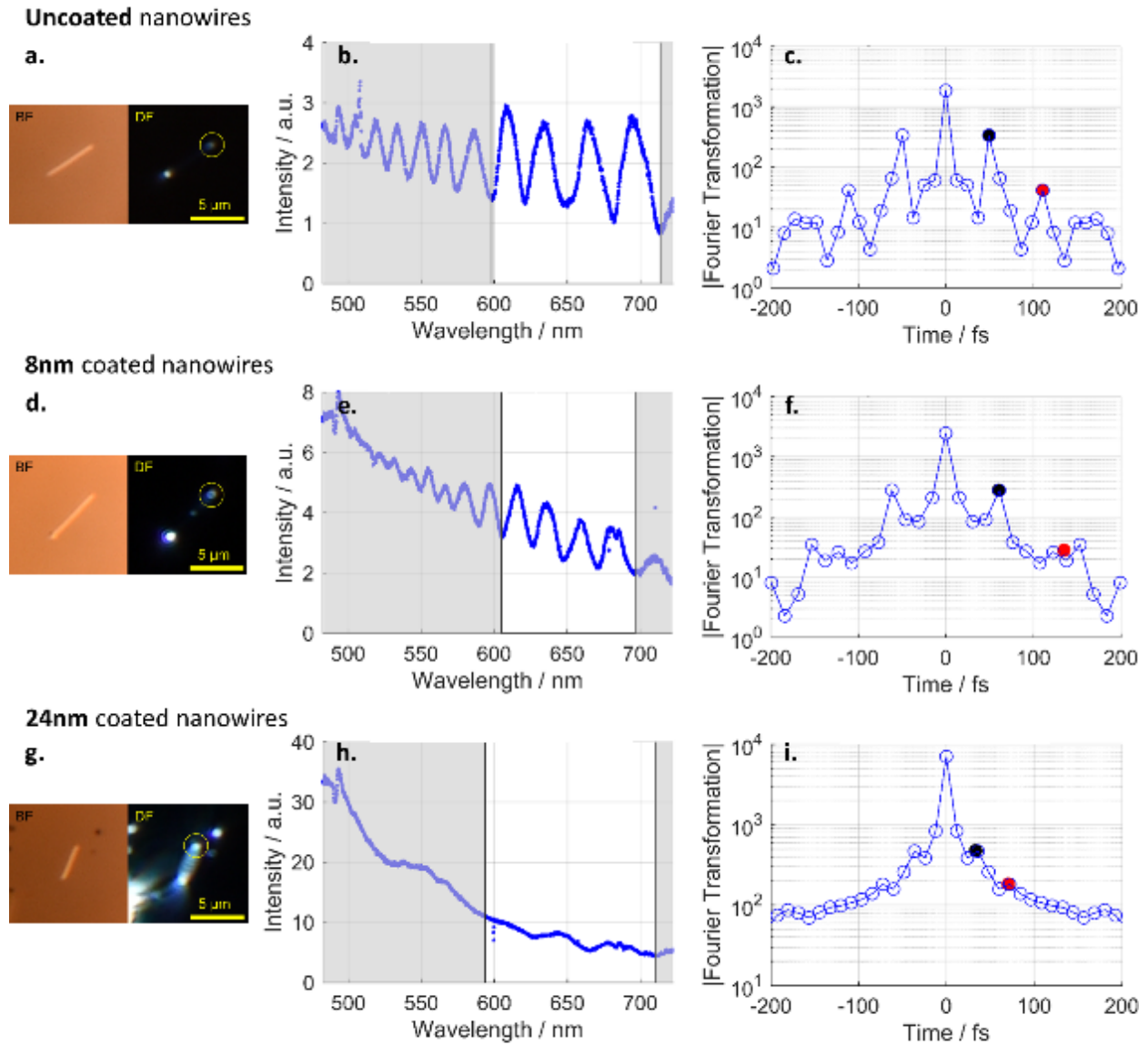

Figure 7. Detailed analysis of scattering spectra. Left column: bright-field (BF) and dark-field (DF) images of single nanowires without coating (a), with 8 nm silica coating (d), and with 24 nm silica coating (g). A pronounced scattering merely at the end facets is observed. Middle column: Measured scattering spectra of light from the front facets revealing the plasmonic Fabry-Pérot-like resonances. Right column: Fourier transform of the spectra (in the frequency

domain) around 650 nm (white area in middle column) revealing the plasmon round trip time (black dot) and the second round trip time at twice the time (red dot).

The Fourier transforms are presented in the right column of Fig. 7 in the range of -200 to 200fs. The full range Fourier transformations are available in the supplementary**.** Note that the resolution of the Fourier transformation is limited by the number of data points and the range of the analyzed spectrum. As already shown by Alione et al.[38] and Hofstetter et al. [36,37], the position of the first peak (black dot) in the sidebands of the Fourier transformation contains directly the information of the plasmon round trip time. For each nanowire the peak position was extracted by fitting the data in the vicinity of the discrete maximum with a parabolic function. The apex position was assigned to be the plasmon round trip time $T_{rt}$ in the nanowire. The length $L$ of the nanowire is measured from the dark-field images. Comparing to the vacuum speed of light, $c$, the real part of the effective refractive index is given by

$$\text{Re}(n_{\text{eff}}) = \frac{c \cdot T_{\text{rt}}}{2L} \tag{1}$$

In order to compare the experimental findings with theory, we performed numerical simulations with a finite element solver (JCMsuite) for a fixed frequency corresponding to a vacuum wavelength of $\lambda_0 = 650$ nm. This wavelength was chosen, as it is centered in the experimentally evaluated range. The field distribution of the fundamental propagating mode (Fig. 6a) was computed together with the corresponding complex effective refractive index $n_{\text{eff}}$. The real part of $n_{\text{eff}}$ contains information about the wavelength of the propagating plasmon polariton ($\lambda_{\text{spp}} = \text{Re}(n_{\text{eff}}) \cdot \lambda_0$) whereas the imaginary part is related to its damping constant ($\alpha = 4\pi \cdot \text{Im}(n_{\text{eff}})/\lambda_0$). Further, we computed the reflectivity $R$ of end facets of the silver nanowires. To this end, we used the field computed with the propagating mode solver as a source that was launched into a three-dimensional (3D) computational domain with a semi-infinite silver nanowire (Fig. 8b). In this computation, we set all imaginary parts of the contributing permittivities to zero to eliminate any damping due to absorption during propagation. We then compute the ratio of the reflected flux with the launched flux, to derive the reflectivity of a single end facet. In all simulations the cross-section of the silver nanowire was assumed to be a regular pentagon.[51] The considered nanowire is either lying directly on a glass substrate or with an additional coating, i.e., with a pentagonal layer with a refractive index of 1.4 representing the silica. In the 3D reflectivity simulations, we restrict ourselves to flat end facets for simplicity. In simulations without any substrate (isotropic surrounding around the wire and cylindrical symmetry), we found that the shape of the end facet (tapered tip) has almost no impact on its reflectivity. The results for the real and the imaginary part of $n_{\text{eff}}$ are plotted in Fig. 8c and d, respectively.

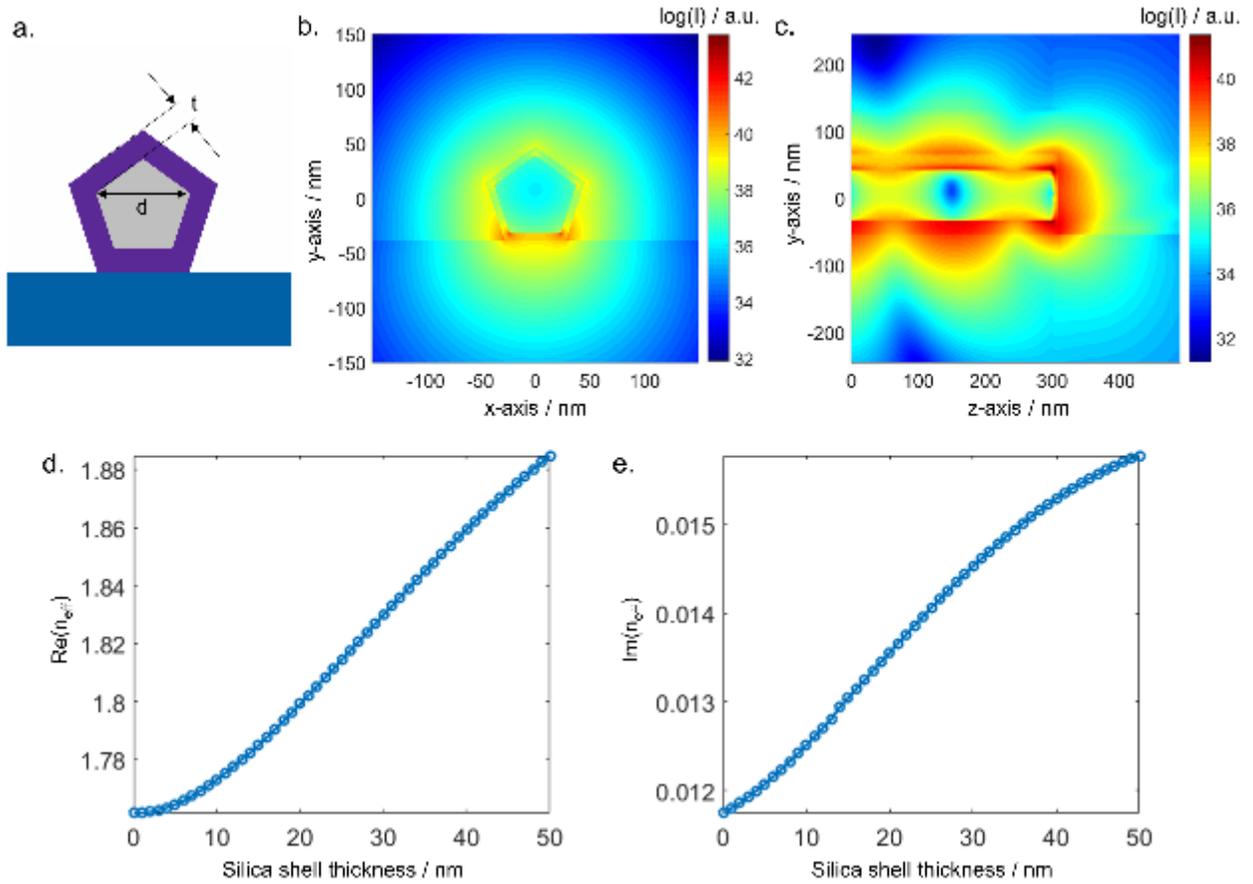

Figure 8. Numerical simulations: (a) Cross section of the silver nanowire of diameter *d* coated with a silica shell of thickness *t*. (b) Finite element propagating mode calculations of the logarithmic intensity distribution of the guided plasmon mode for $\lambda_0=650$ nm. (c) Logarithmic intensity of a full 3D computation (cut through the center of the wire). The propagating mode is used as a source to compute the reflectivity at a nanowire end facet. For this computation, the imaginary parts of silver have been set to zero. The computations further yield (d) the real part and (e) the imaginary part of the effective refractive index of the guided mode depending on the shell thickness.

Figure 9 summarizes our experimental findings with a comparison to theoretical predictions. Figure 9a shows a scatter plot of the measured Re($n_{eff}$) as a function of the corresponding nanowire length *L* of the investigated specimen together with the values found in our numerical simulations (indicated by straight horizontal lines as $n_{eff}$ is independent of the waveguide length). We find that relatively short nanowires show a larger variance of Re($n_{eff}$) which we attribute to general limitations of the applied dark-field microscopy. Nanowires with short lengths close to the diffraction limit suffer from three uncertainties: First, the relative error of the measured nanowire lengths is large for short nanowires. Second, the front facet and end facet are spatially not well separated so that contributions from both ends might enter the detection. Further, the free spectral range of short nanowires is relatively large so that only a few oscillations are apparent in the scattering spectrum. With respect to our analysis based on a Fourier transformation this means that the central peak will be very close to the peak representing one roundtrip time, which is thus hard to resolve.

However, we find that for nanowires longer than 3 µm (white area in Fig. 9a), the method becomes reliable. The variance reduces significantly and the experimentally derived Re($n_{eff}$) scatter around the theoretically expected value (dashed line Fig. 9a). We further included the mean and standard deviation of Re($n_{eff}$) for the nanowires longer than 3 µm as solid lines and as colored shaded areas for the coated and uncoated nanowires, respectively. The simulated values are covered by the standard deviations. From this we conclude that the silica shell has only minor influence on Re($n_{eff}$) as already predicted by our simulations.

More importantly, the plasmon damping can be extracted from the data. An additional peak is expected in the Fourier spectrum at twice the round trip time ($t=2T_{rt}$).[36-38] The harmonic peak amplitude ratio, i.e., the ratio of the second peak $p_2$ and the first peak $p_1$, is a direct measure of the overall round trip losses *RTL*. These losses stem from imperfect reflectivities $R_1$ and $R_2$ of the end facets and the damping constant α of the SPPs. Besides sole Ohmic damping of the SPP, the apparent α may contain losses by scattering, e.g., due to roughness of the metal-dielectric interface of the AgNWs or imperfections of the silica coating. Reflectivities and damping are connected to the harmonic amplitude ratio by the survival probability $P_{survive}$

$$P_{survive} = 1 - \text{RTL} = \left(\frac{p_2}{p_1}\right)^2 = (R \cdot e^{-\alpha L})^2 \qquad (2)$$

i.e., the ratio of SPPs surviving one round trip. Here we assume equal reflectivities at both facets ($R = \sqrt{R_1 R_2}$). The relatively high round trip losses correspond to small peak heights $p_2$ in the Fourier spectra as can be seen in Fig. 7 where the black dot marks $p_1$ and the red dot $p_2$ (on a log scale). The offset, overlying tilt, and noise in the scattering spectrum leads to a background in the Fourier spectra, while SPP damping and dispersion leads to a broadening of the peaks. Thus, the peak height $p_2$ could not be extracted for all nanowires, but we only deduced an upper limit for the harmonic amplitude ratio or $P_{survive}$, respectively. In Fig. 9b, we show $P_{survive}$ versus the corresponding length of the nanowire specimen. Solid points represent data, where a second peak could be clearly identified, while hollow points indicate situations, where only an upper limit could be estimated. For the nanowires without coating, we find the longest specimen. Half of investigated specimens (9 out of 19) of length greater than 3µm show a pronounced second peak in the Fourier spectra (solid blue points). For the nanowires with silica coating of 8 nm, 3 out of 13 Fourier spectra show a pronounced peak height $p_2$ (solid red boxes in Fig. 9b). For these specimens $P_{survive}$ is comparable to the nanowires without coating. For the remaining ten spectra, we again interpret the results as upper bounds (hollow red boxes in Fig. 9b). Additionally, we were able to extract $P_{survive}$ of a single nanowire with 24 nm silica coating. Together with this scattered data, we plot the theoretically expected $P_{survive}$ derived from the numerical simulations. These findings suggest that the silica coating is doing no harm to the bare AgNWs and the plasmonic performance. Though, we had to sonicate coated AgNWs for longer times to avoid agglomeration, which led to more broken, and thus on average shorter nanowires. However, we believe that this problem can be avoided, e.g., by more diluted samples that can be spin-casted multiple times to achieve equal densities on substrates. For the sample with the thickest coatings we observed secondary silica particles, partly attached to the AgNWs, which limits the method somehow as these secondary particles act as efficient scatterers of the SPP field representing additional loss channels. Nevertheless, we believe that again the synthesis procedure could be optimized to avoid such secondary particles more efficiently.

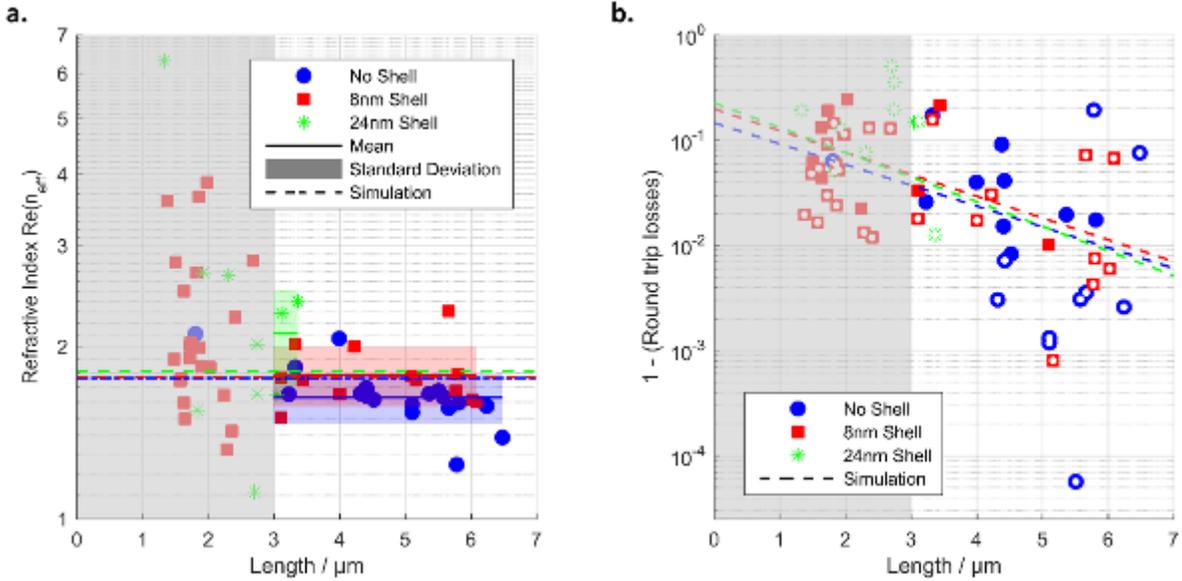

Figure 9. (a) Real parts of the effective refractive index Re($n_{eff}$) as experimentally derived (scattered markers) and theoretical predicted (dashed lines). Solid lines and shaded areas (in green, red and blue) are mean and standard deviation of each group of nanowires (without shell, with 8nm shell, with 24nm shell), respectively. Note that our method becomes reliable (see text) for nanowires longer than 3µm (white area). (b) Survival probability $P_{survival}$ as extracted from Fourier spectra for the nanowire specimens (markers). Full markers represent data where a clear second round trip peak could be observed, while hollow markers represent data where only an upper limit could be deduced. The dashed lines show the results of the numerical calculation.

Finally, our findings are also nicely supported by the evaluation of the EELS spectra. Firstly, the EELS data clearly show the standing wave pattern expected for the plasmonic Fabry-Pérot resonators. This further supports our understanding that we indeed probe the Fabry-Pérot resonances with our optical studies and do not observe oscillating spectra that are just due to accidentally formed micro-resonators[34]. Here, the EELS spectra could be evaluated best in the infrared spectral range, i.e., for relatively small electron energy loss. In Fig. 6, several harmonics are clearly visible for different electron loss energies for both a coated and an uncoated nanowire. Applying the same Fourier algorithm to the center-line spectrum (Fig 6b and 6e, horizontal line) we again create a Fourier spectrum (Fig. 10a and 10b) for which we apply the same analysis as for the optical investigations. The real part of the effective refractive index were found to be 1.70 and 1.98, for the uncoated and coated AgNWs, respectively. This is in agreement to simulated values of 1.59 and 1.82. The position of the second peak is indicated by the red dot. $P_{survival}$ is estimated to be 0.039 and 0.092 from this. These values also match the simulated results of 0.043 and 0.095, calculated from the imaginary part of the refractive index of 0.011 and 0.017 and reflectivities of 0.34 and 0.46, respectively. The simulated first and second round trip time is shown in Fig. 10 as black and red vertical line, respectively. The simulated damping results in an expectation of the second peak height (horizontal red line). Both matching peak position and height illustrates the excellent agreement between our measurements and simulations. This supports the conclusion that especially the damping of the SPPs is barely affected by the silica coating.

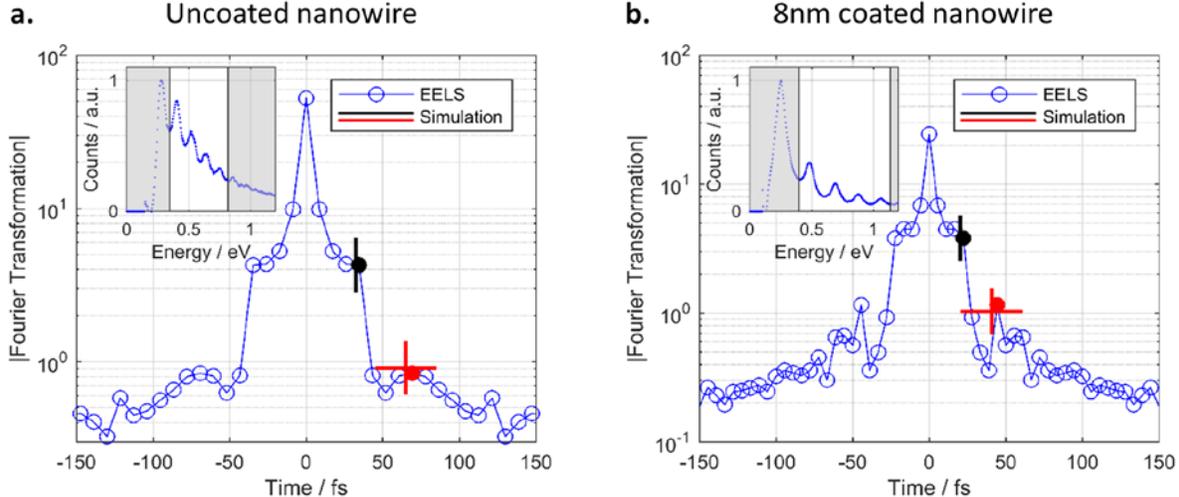

Figure 10. Detailed analysis of the EELS spectra of (a) the uncoated and (b) the 8 nm coated AgNW. The Fourier transformations reveal the expected peaks at the first round trip time (black dot) and the second round trip time (red dot). Both position and peak amplitude ratio match the simulation (black and red lines). The inset indicates the analyzed part of the EELS spectra (white area).

## Conclusion

We reported on the synthesis of thin and short silver nanowires (70 nm in diameter and few µm in length) on which we apply a modified Stöber method to create an optimized silica coating (here 8 nm and 24 nm thick). Such coatings enable functionalization with emitters and distance control. We use transmission electron microscopy, single-particle dark-field scattering spectroscopy and thorough numerical calculations to demonstrate that the modified Stöber method yields well-defined smooth silica coatings that do no harm to the plasmonic properties of the silver nanowires.

We conclude that our modified synthesis method opens the door for the construction of well-defined coupled systems of quantum emitters and plasmonic resonators and their usage in integrated optical devices[19-21], i.e. all-optical transistors[52,53] or active plasmonic sensors[54]. Such coatings have been frequently used for gold nanoparticles so far but are now accessible to silver nanostructures like nanowires, too. We thus expect our work to have a broad impact on future studies on hybrid plasmonic systems based on silver nanostructures.

With respect to surface plasmon lasing[13-15] or strong coupling experiments[8,12], we believe that the geometric advantage of nanowire modes to be able to overlap spatially with much more gain medium (or emitters) compared to nanoparticle modes can now be accessed much more conveniently as the silica coating allows for incorporation of emitters[55].

# Methods

**Materials.** Silver nitrate (99%), ethylene glycol (99.8%), poly(vinyl pyrrolidone) (Mw= 40000), tetraethyl orthosilicate (TEOS), sodium hydroxide, sodium chloride, ammonia solution (28 wt%), absolute ethanol were purchased from Aldrich. All the chemicals were used without further purification. Cover glass slips were cleaned with Hellmanex solution.

**Synthesis of silver nanowires.** The synthesis of silver nanowires followed the method develop by Robert K. Y. Li.[39] In a typical run, 2.005 g poly(vinyl pyrrolidone) (Mw= 40000) was dissolved in 34 ml ethylene glycol under heating and stirring. After the temperature was stable at 160 °C, 40 µl NaCl solution (0.8 mM) was added into the flask. One minute later, 6 ml AgNO$_3$ (0.1 M) in ethylene glycol solution was injected by a syringe pump at a flow rate of 10 µl s$^{-1}$. Once the reaction solution turned turbid, all residual AgNO$_3$ solution was added into the flask immediately. The mixture was then continuously stirred for 30 min at 160 °C followed by natural cooling down to ambient temperature. The as-prepared AgNWs were collected via centrifugation and stored in ethanol at 4 °C.

**Silica coating of silver nanowires.** For the silica coating, 35 µl AgNWs solution (6 mg ml$^{-1}$) was dispersed in the mixture of 6 ml absolute ethanol, 2 ml deionized water and 200 µl NaOH (0.1 M) under magnetic stirring. To gain silver nanowires with 8 nm silica shell, 1 ml TEOS solution (0.26 vol% of TEOS in ethanol) was added slowly with continuous stirring for 15 min. The mixture was stirred at 500 rpm for further 3h at ambient temperature. As-prepared silica coated silver nanowires (noted as SiO$_2$8nm@AgNWs) were collected via centrifugation after rinsed in water. Silver nanowires with around 25 nm silica shell (noted as SiO$_2$24nm@AgNWs) were achieved by tuning the concentration of TEOS to 0.52 vol% keeping the other conditions the same. As a comparison, 200 µl NaOH (0.1 M) were replaced by 200 µl ammonia solution (28 wt%) under the same condition. Samples of silica coated silver nanowires with around 8 nm and 24 nm shell are noted as A-SiO$_2$8nm@AgNWs and A-SiO$_2$24nm@AgNWs correspondingly.

**Characterization.** The morphology of the nanowires was investigated by transmission electron microscopy (TEM) using a JEOL JEM-2100 (JEOL GmbH, Eching, Germany) at an acceleration voltage of 200 kV. Samples were prepared by placing a drop of the diluted nanowires solution on the carbon-coated copper grids and dried under ambient temperature. The ultraviolet-visible spectra (UV-VIS spectra) were measured by Lambda 650 spectrometer supplied by Perkin-Elmer or Agilent 8453.

**Bright-field and dark-field optical microscopy.** The plasmonic modes of individual nanowires were analyzed by white-light scattering in a home-build microscope setup. Samples were prepared on cleaned cover-glass slips from nanowire dispersions which were diluted with ethanol and placed into an ultrasonic bath for 5 to 15 minutes. Subsequently, 50 to 150µl of the dispersion were spin-coated onto a clean cover glass slip. This procedure was optimized for every dispersion to achieve a convenient distribution of single nanowires on the cover glass.

Single nanowires were imaged by bright-field and dark-field illumination using a Nikon D5500 camera. A Xenon arc lamp equipped with a linear polarizer was used for illumination.
Scattered light was analyzed in a confocal setup with an Andor Newton DU-920P-BV CCD-camera mounted on an Acton Spectra Pro 2500i spectrometer.

**Electron energy-loss spectroscopy.** Measurements of the plasmonic response were conducted in the Zeiss SESAM microscope operated at a high tension of 200 kV. Using an electrostatic electron monochromator, this instrument offers an energy resolution below 100 meV. EELS spectra were recorded at a dispersion of 0.0047 eV/pixel using the in-column MANDOLINE energy filter. 100 or 400 spectra were collected along the edge of the wire with a dwell time of typically 0.5 s per spectrum. The electron beam was always outside the wire (aloof geometry) to minimize volume contributions to the spectra. High-angle annular dark-field (HAADF) images were acquired for scattering angles ranging from 28 to 190 mrad.

## Acknowledgement
Funding by the Deutsche Forschungsgemeinschaft DFG is acknowledged through the Collaborative Research Centre 951 'Hybrid Inorganic/Organic Systems for Optoelectronics (HIOS)' within project B2.

## Competing interests
The authors declare no competing interests.

## Data availability statement
All data generated or analyzed during this study are included in this published article (and its Supplementary Information files).

## Contributions
AgNW synthesis was developed by Y.Z. and Y.L. Y.Z. performed chemical synthesis. Y.Z. and Z.K. performed TEM measurements. Optical measurements were conducted and analyzed by M.R. Numerical simulations were performed by G.K. Electron energy-loss spectroscopy was performed by W.S. and P.v.A. O.B. and Y.L. supervised the study. All authors contributed to data interpretation and manuscript revision.

## Corresponding author
Correspondence to Martin Rothe.

# Supporting Information

# Silver nanowires with optimized silica coating as versatile plasmonic resonators

*Martin Rothe,[1] Yuhang Zhao,[2] Günter Kewes,[1] Zdravko Kochovski,[2] Wilfried Sigle,[3] Peter A. van Aken,[3] Christoph Koch,[4] Matthias Ballauff,[2,5] Yan Lu,[2,6] and Oliver Benson[1]*

[1] *Humboldt Universität zu Berlin & IRIS Adlershof, Nanooptics, Newtonstraße 15, 12489 Berlin, Germany*
[2] *Helmholtz Zentrum Berlin für Materialien und Energie, Institute of Soft Matter and Functional Materials, Hahn-Meitner-Platz 1, 14109 Berlin, Germany*
[3] *Stuttgart Center for Electron Microscopy, Max Planck Institute for Solid State Research, Heisenbergstr. 1, 70569 Stuttgart, Germany*
[4] *Humboldt Universität zu Berlin & IRIS Adlershof, Structure Research and Electron Microscopy, Newtonstraße 15, 12489 Berlin, Germany*
[5] *Humboldt Universität zu Berlin, Department of Physics, 12489 Berlin, Germany*
[6] *Institute of Chemistry, University of Potsdam, 14467 Potsdam, Germany*

The purpose of this supplement is to present the results of the optical measurements for each single nanowire.

Fig. S1 to S2 present the bright-field and the dark-field images of each single pure silver nanowire investigated. The scattering from the end facets of the nanowires is clearly visible in the dark-field images. For further analysis, the scattered light from the front facet of the nanowires, i.e. the upper right end in the images, was studied spectroscopically. The measured spectra were background subtracted and normalized with the spectrum of the white light source. The resulting spectra are presented in the left column of fig. S3-S6. The white area around 650 nm was further investigated by transforming this part to frequency space and subsequent Fourier transformation. The result is presented in the center column. The right column presents a zoom from -200 fs to +200fs where peaks are expected for every multiple of the plasmon round trip time. The first peak, labeled by a black spot indicates the time for one plasmon round trip. At twice this time a second peak is expected. This position or a peak, if visible, is marked with a red spot. If a second peak is apparent the plasmon losses have been calculated. If not, the noise level was interpreted as upper limit for the second peak indicating a lower limit for the losses.

In analogy, Fig. S7 to S17 present the same results for silver nanowires coated with 8nm silica shell. The amount of samples has been increased to cover the greater distribution of lengths.

Fig. S18 to S20 show the ten studied silver nanowires with 24nm silica coating. Longitudinal modes are still visible but further effects reduce the quality of the evaluation method.

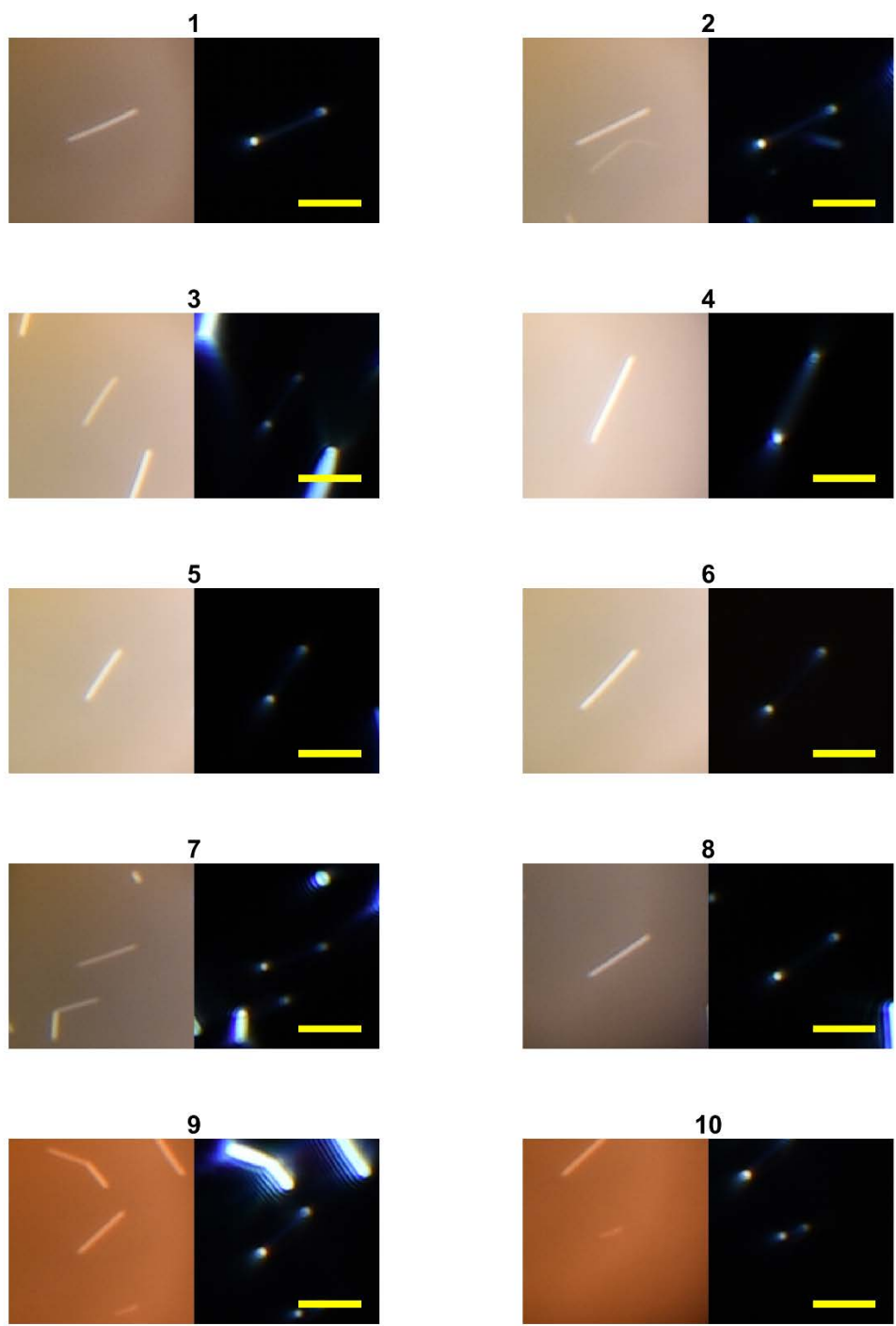

Figure S1. Pure silver nanowire samples 1 to 10 on coverglass are detected by bright-field imaging (left) and dark-field imaging (right), respectively. The scale bar is 5 µm.

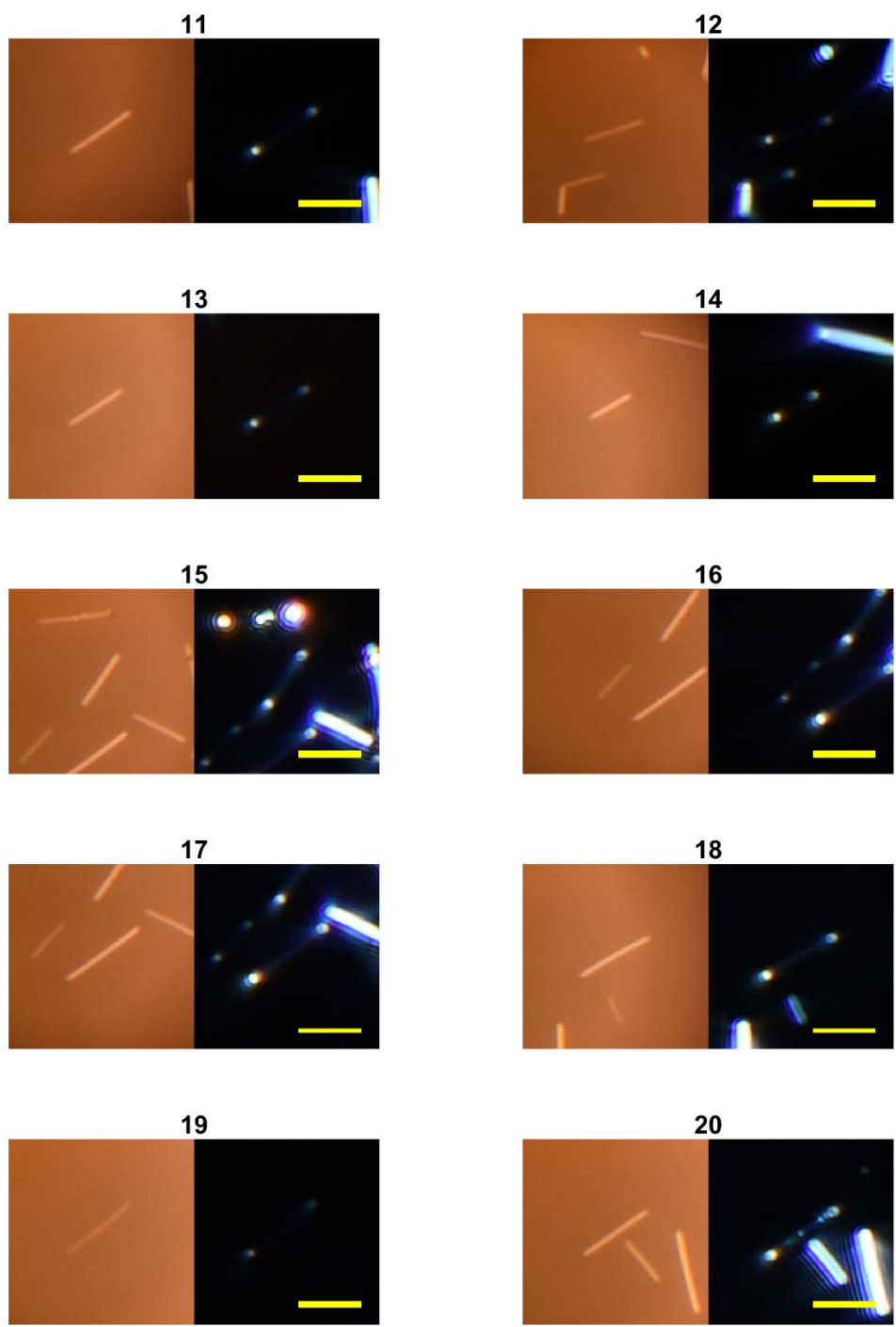

Figure S2. Pure silver nanowire samples 11 to 20 on coverglass are detected by bright-field imaging (left) and dark-field imaging (right), respectively. The scale bar is 5 µm.

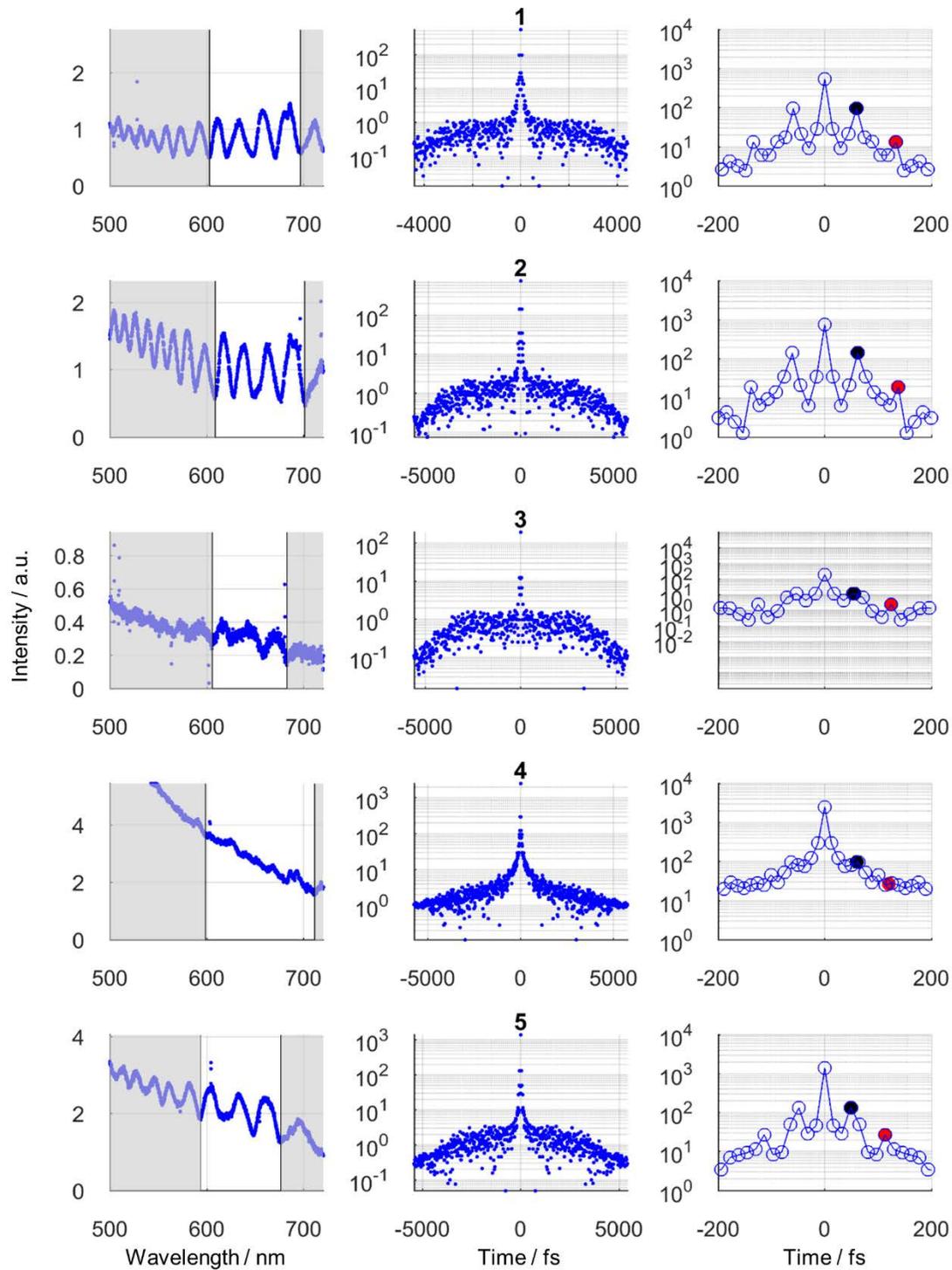

Figure S3. The normalized scattering spectrum (left) of the front facet of pure silver nanowire samples 1 to 5 has been cropped around 650 nm (left, white area) and Fourier transformed (middle) indicating the first (right, black dot) and the second (right, red dot) plasmon round trip.

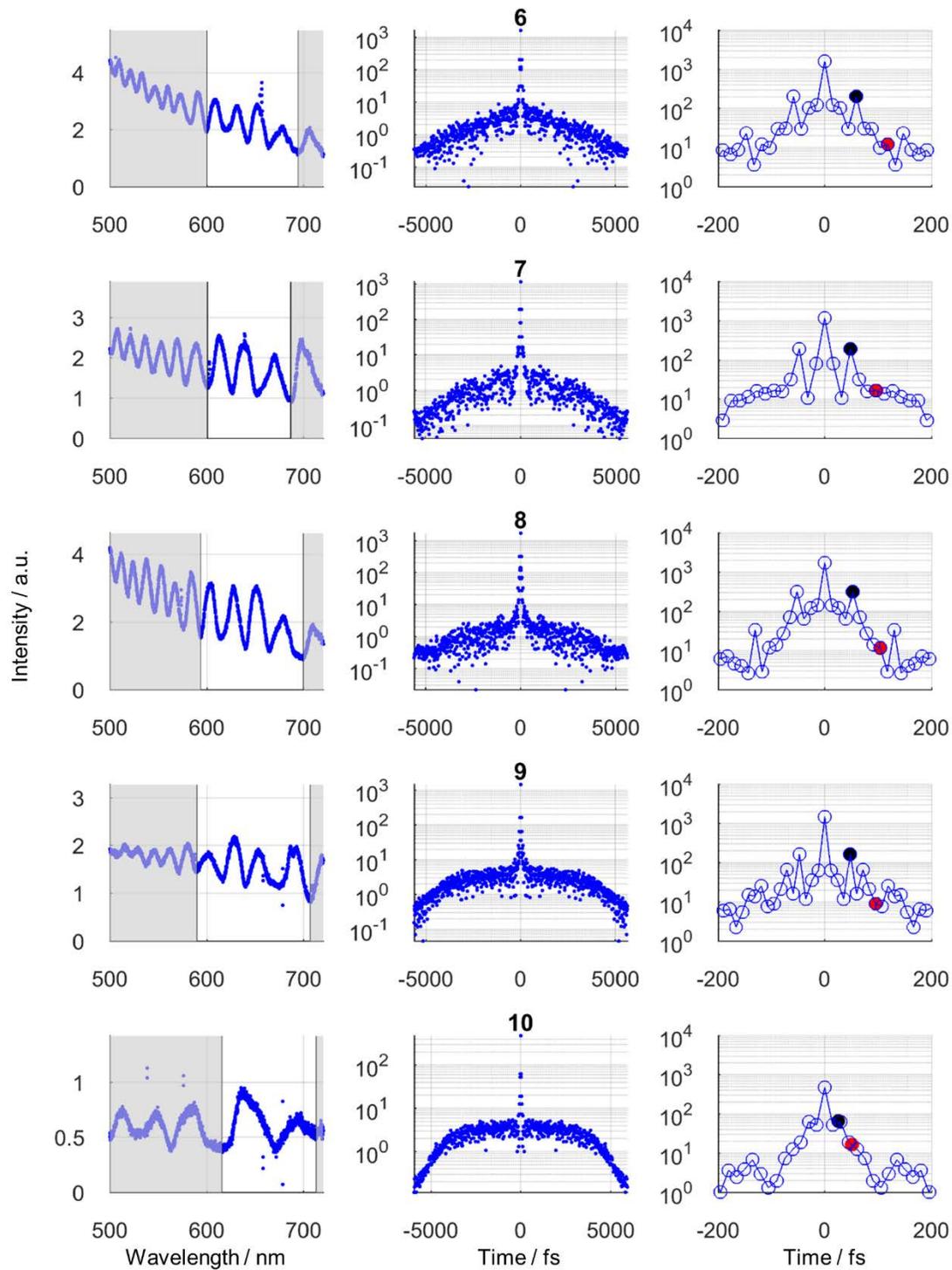

Figure S4. The normalized scattering spectrum (left) of the front facet of pure silver nanowire samples 6 to 10 has been cropped around 650 nm (left, white area) and Fourier transformed (middle) indicating the first (right, black dot) and the second (right, red dot) plasmon round trip.

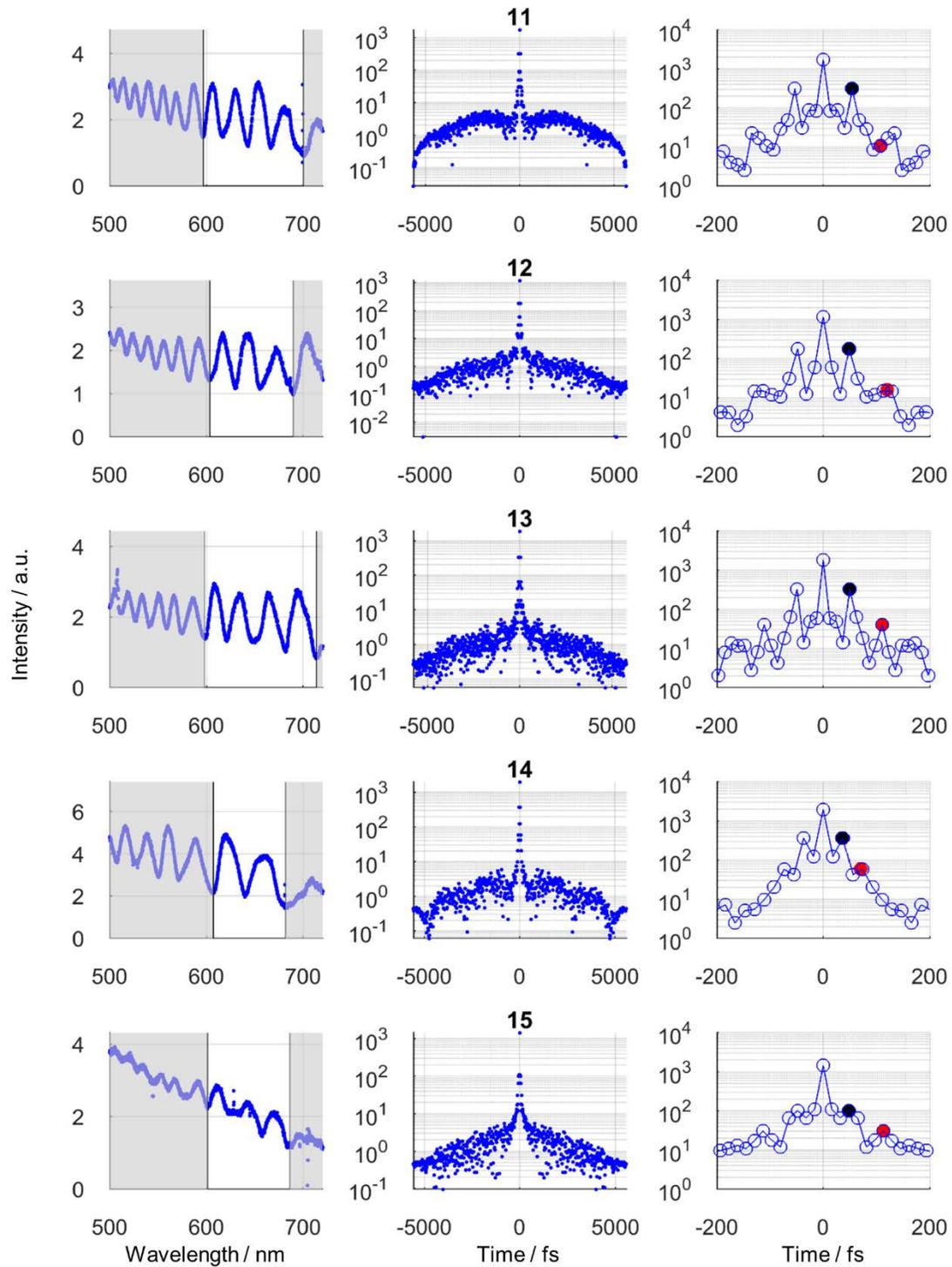

Figure S5. The normalized scattering spectrum (left) of the front facet of pure silver nanowire samples 11 to 15 has been cropped around 650 nm (left, white area) and Fourier transformed (middle) indicating the first (right, black dot) and the second (right, red dot) plasmon round trip.

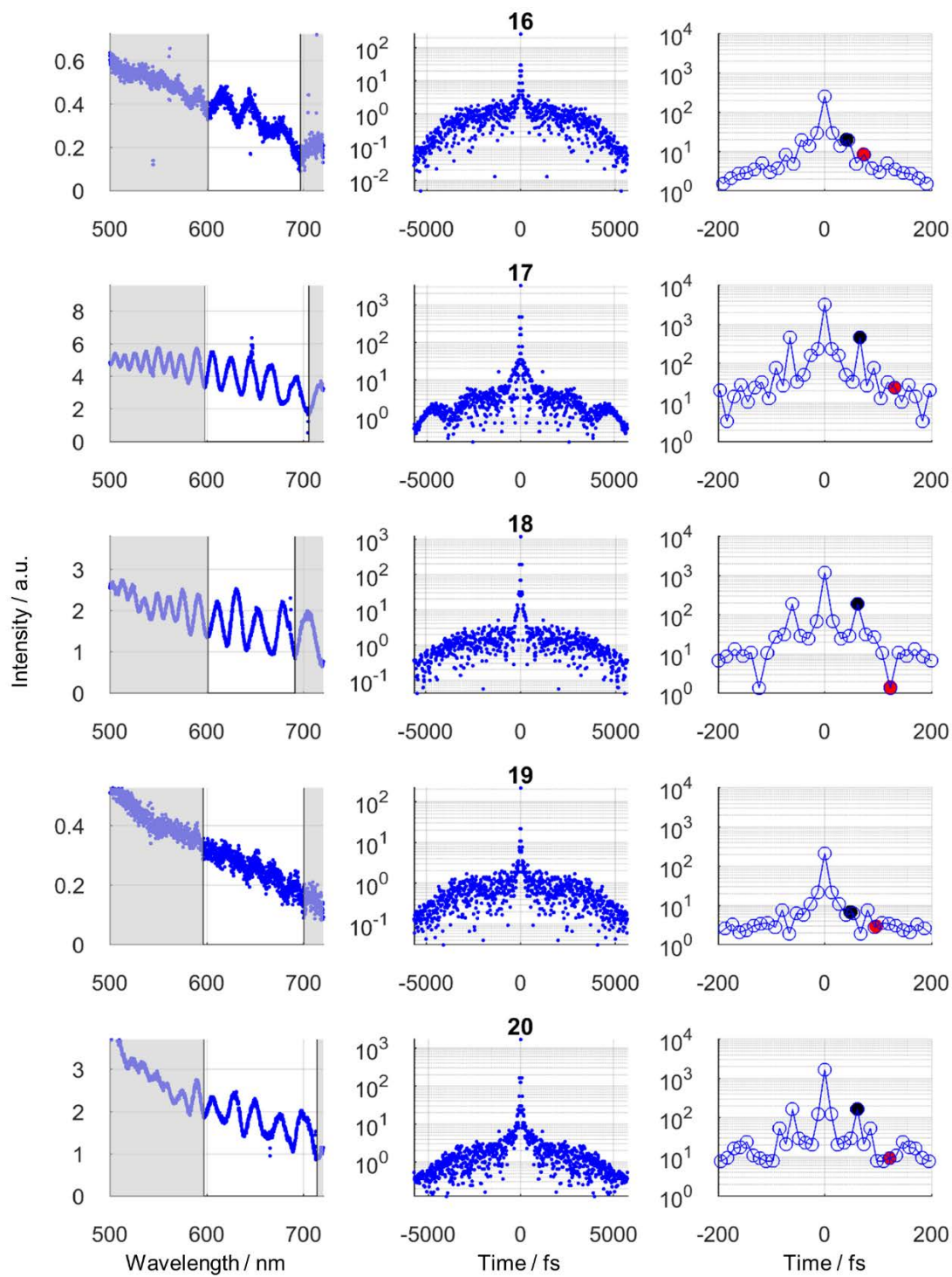

Figure S6. The normalized scattering spectrum (left) of the front facet of pure silver nanowire samples 16 to 20 has been cropped around 650 nm (left, white area) and Fourier transformed (middle) indicating the first (right, black dot) and the second (right, red dot) plasmon round trip.

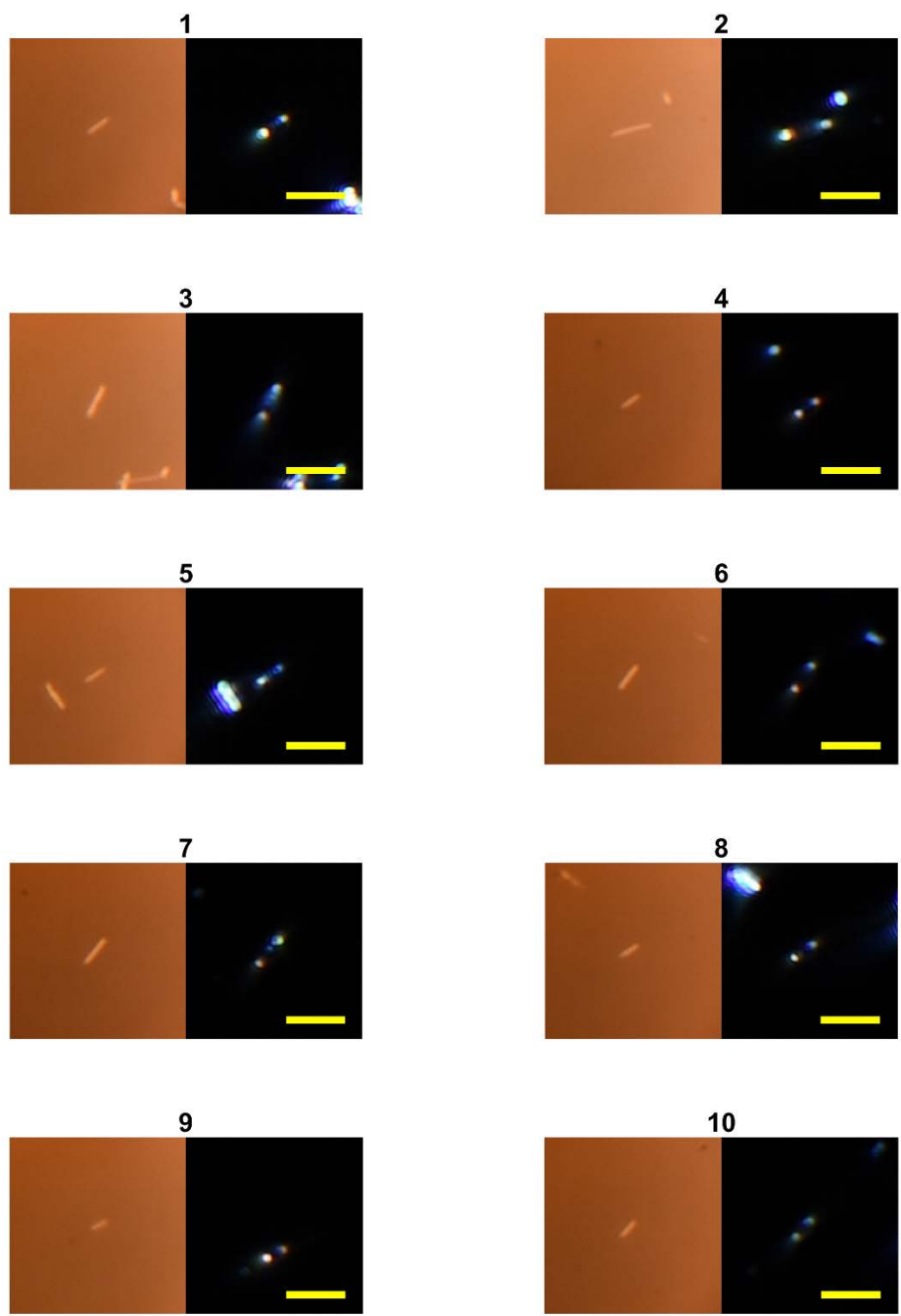

Figure S7. In analogy, silver nanowire with 8 nm silica coating samples 1 to 10 in bright-field imaging (left) and dark-field imaging (right), respectively. The scale bar is 5 µm.

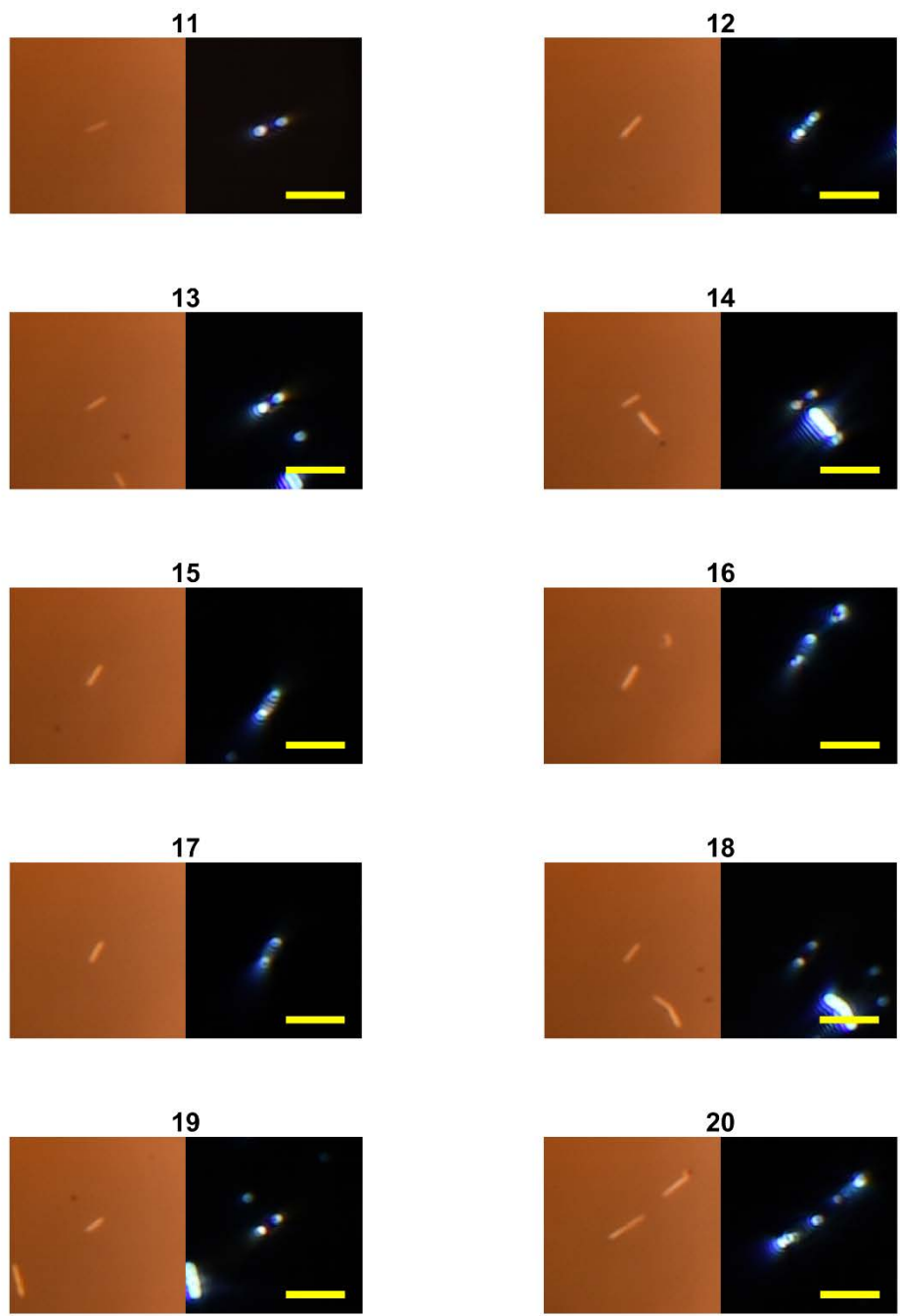

Figure S8. In analogy, silver nanowire with 8 nm silica coating samples 11 to 20 in bright-field imaging (left) and dark-field imaging (right), respectively. The scale bar is 5 µm.

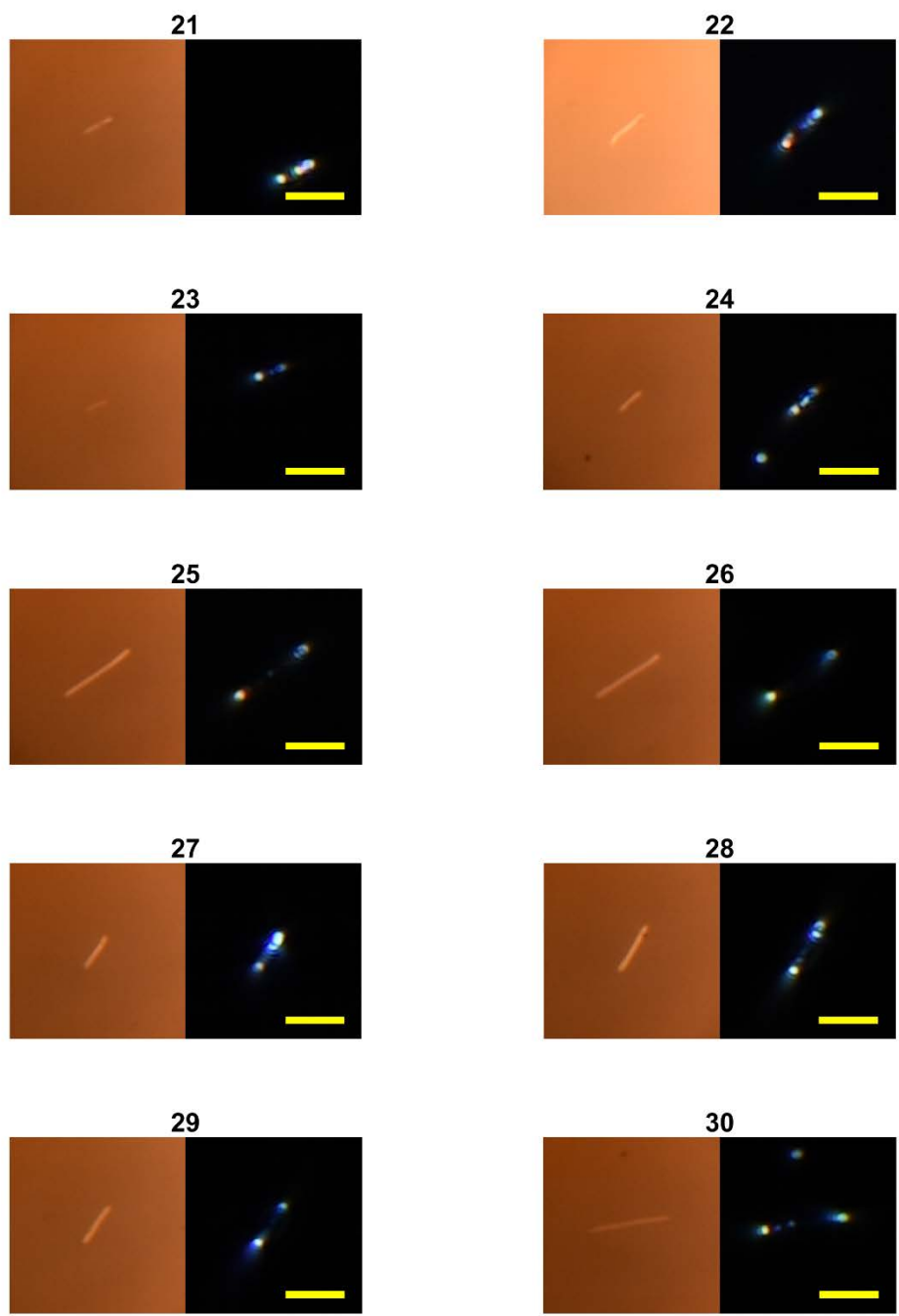

Figure S9. In analogy, silver nanowire with 8 nm silica coating samples 21 to 30 in bright-field imaging (left) and dark-field imaging (right), respectively. The scale bar is 5 µm.

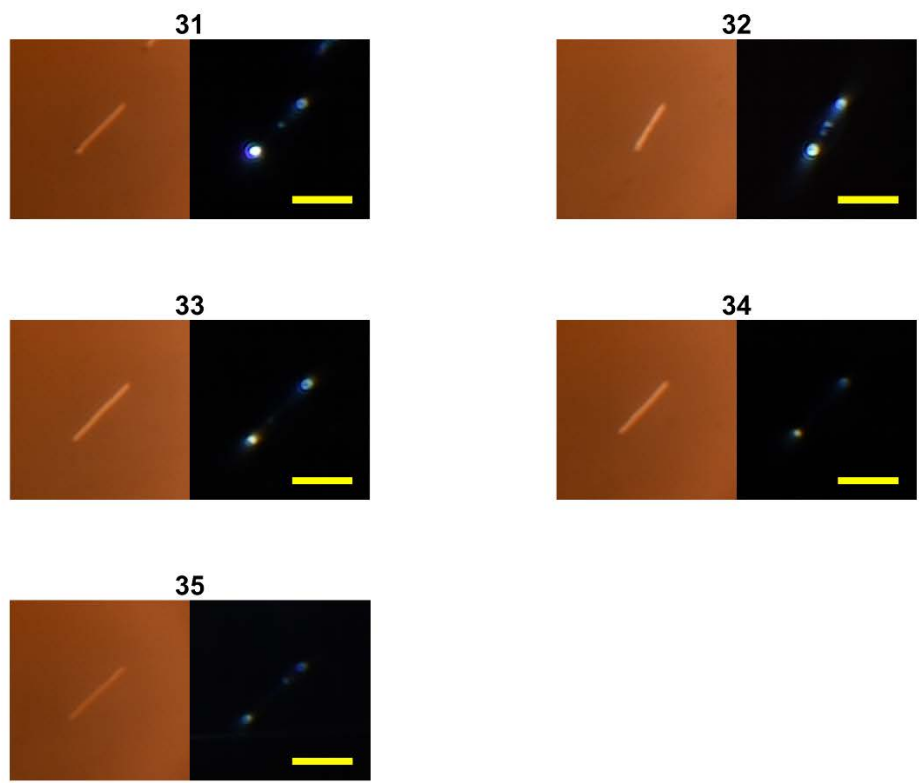

Figure S10. In analogy, silver nanowire with 8 nm silica coating samples 31 to 35 in bright-field imaging (left) and dark-field imaging (right), respectively. The scale bar is 5 µm.

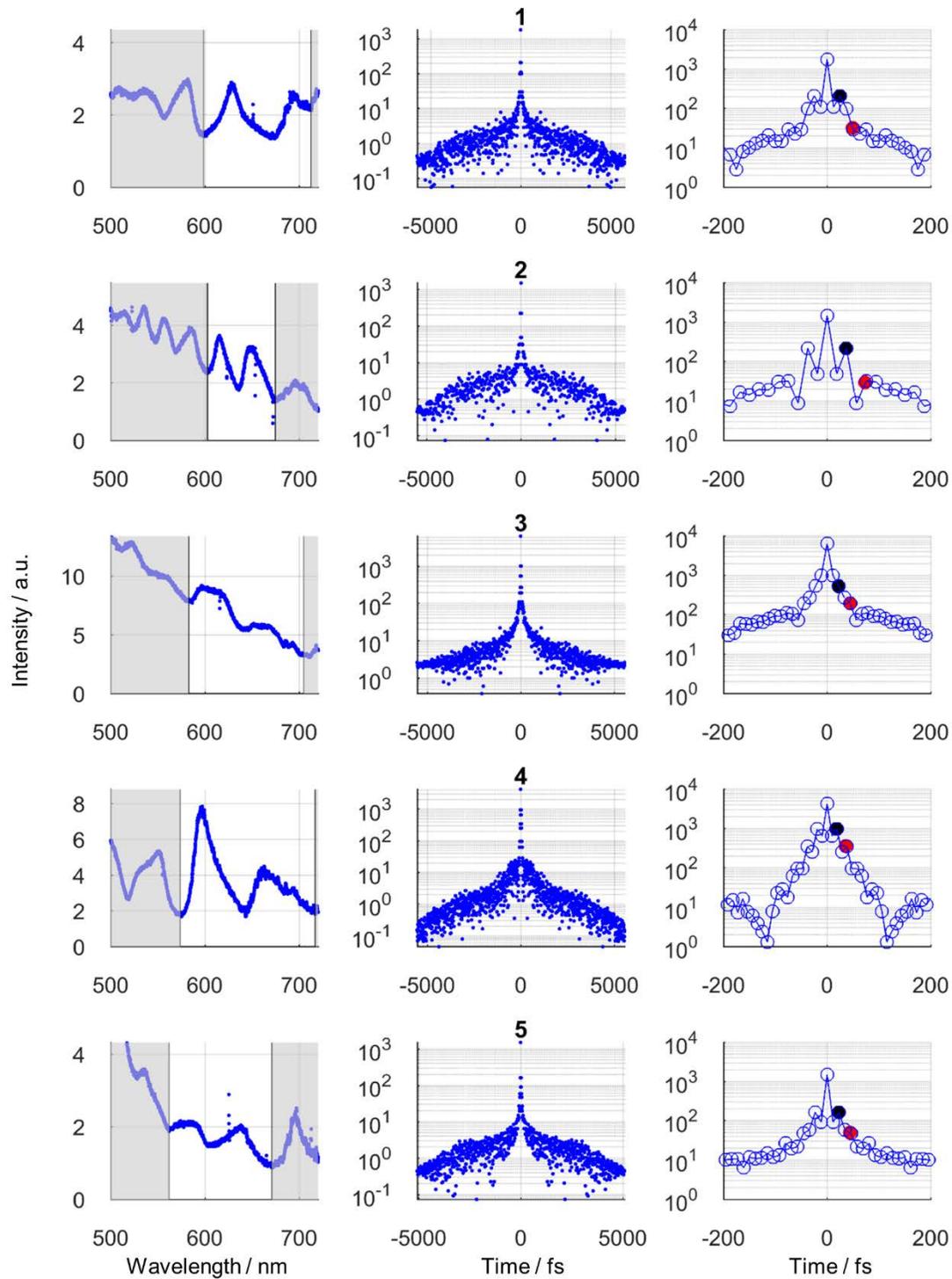

Figure S11. In analogy, the normalized scattering spectrum (left) of silver nanowire with 8 nm silica coating samples 1 to 5, Fourier transformation (middle) and indicated first (right, black dot) and second (right, red dot) plasmon round trip.

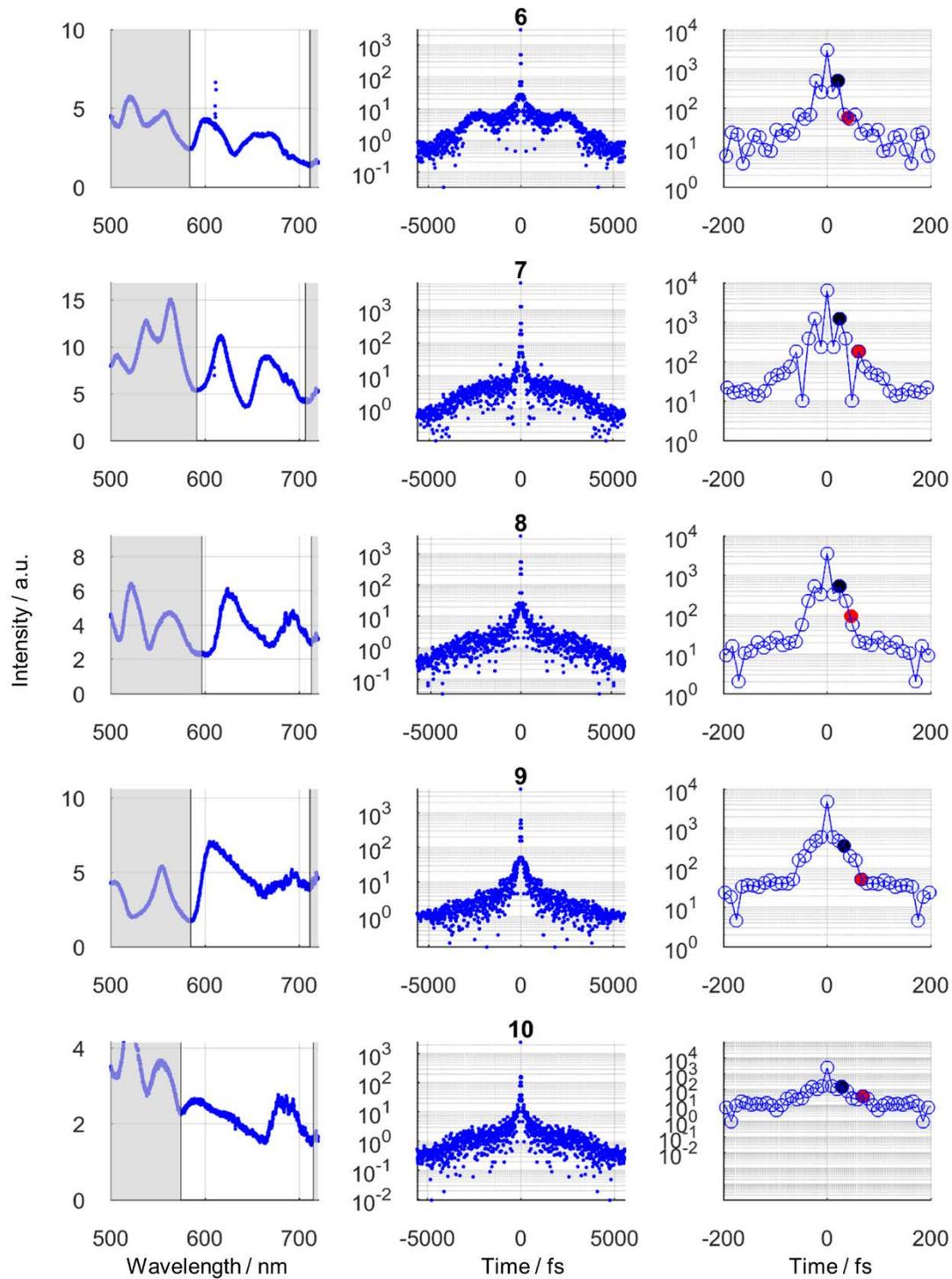

Figure S12. In analogy, the normalized scattering spectrum (left) of silver nanowire with 8 nm silica coating samples 6 to 10, Fourier transformation (middle) and indicated first (right, black dot) and second (right, red dot) plasmon round trip.

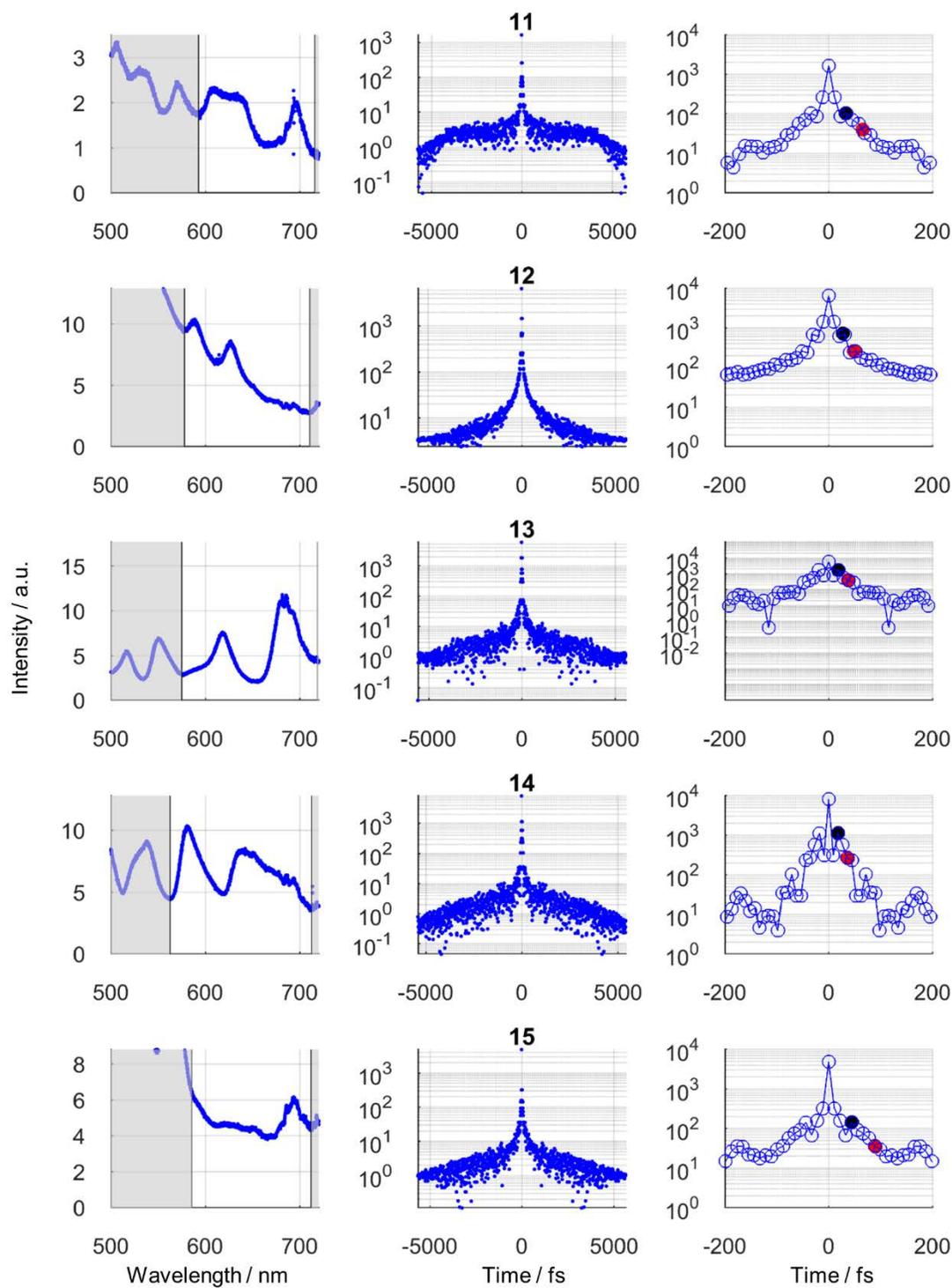

Figure S13. In analogy, the normalized scattering spectrum (left) of silver nanowire with 8 nm silica coating samples 11 to 15, Fourier transformation (middle) and indicated first (right, black dot) and second (right, red dot) plasmon round trip.

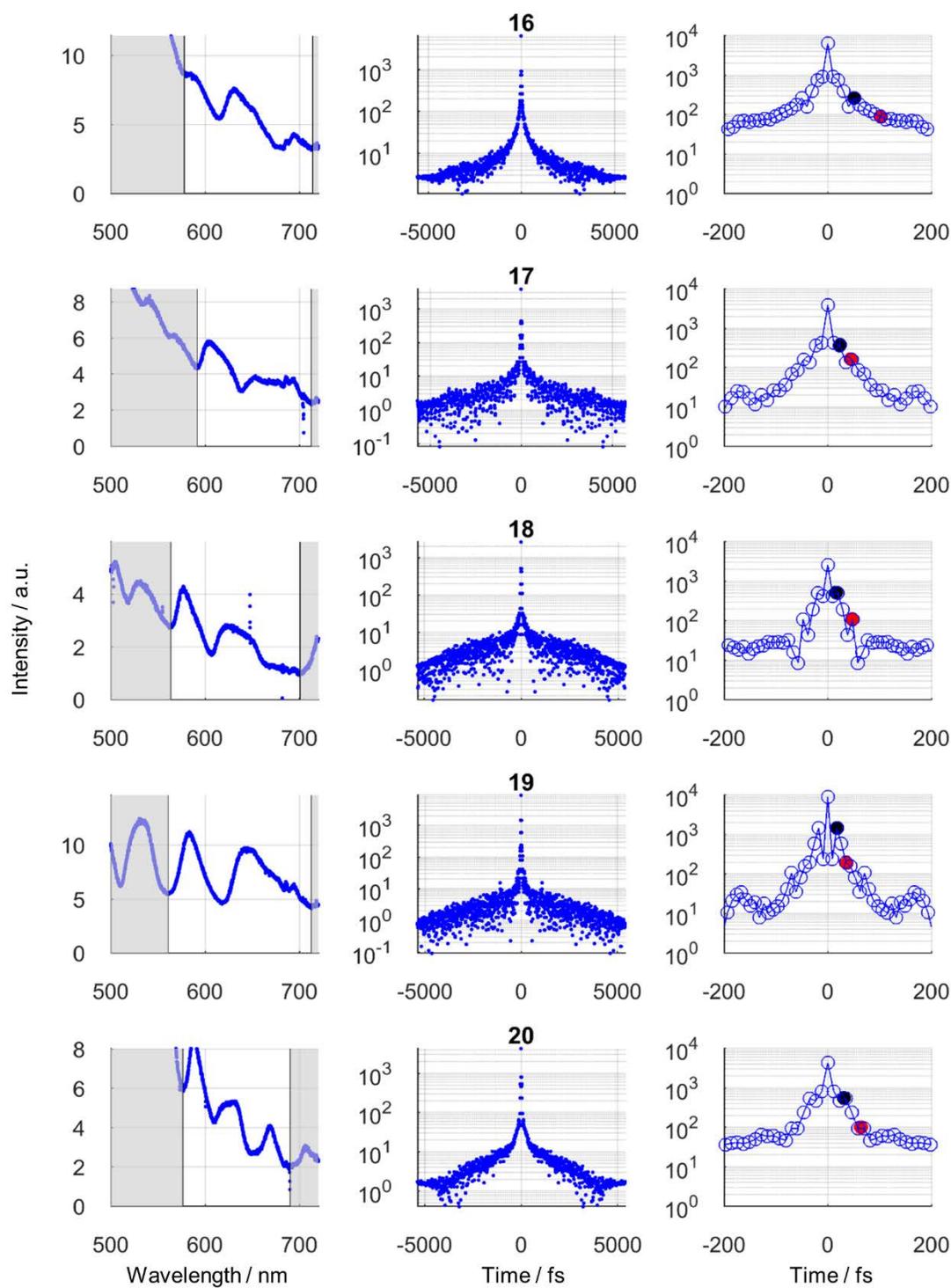

Figure S14. In analogy, the normalized scattering spectrum (left) of silver nanowire with 8 nm silica coating samples 16 to 20, Fourier transformation (middle) and indicated first (right, black dot) and second (right, red dot) plasmon round trip.

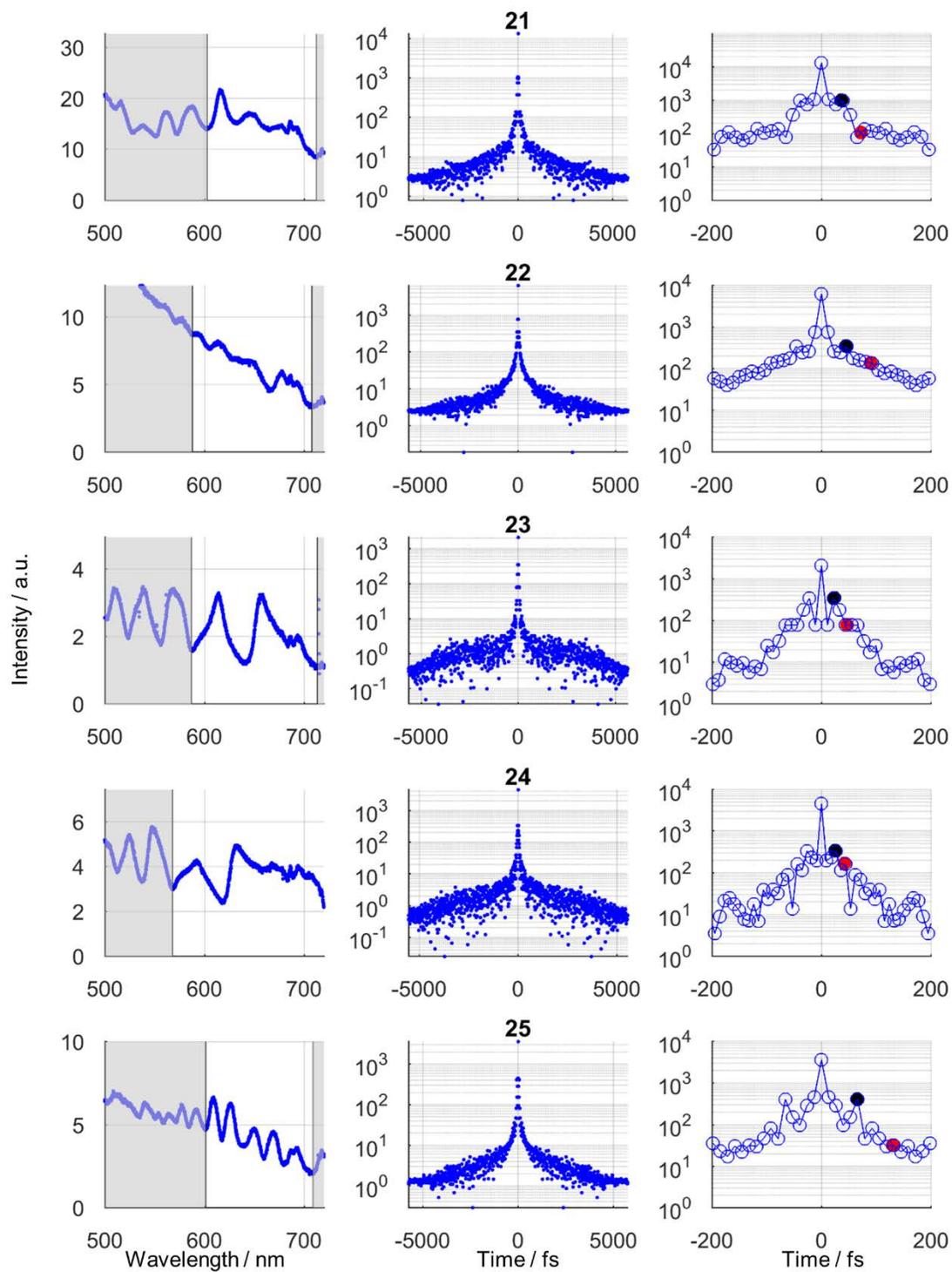

Figure S15. In analogy, the normalized scattering spectrum (left) of silver nanowire with 8 nm silica coating samples 21 to 25, Fourier transformation (middle) and indicated first (right, black dot) and second (right, red dot) plasmon round trip.

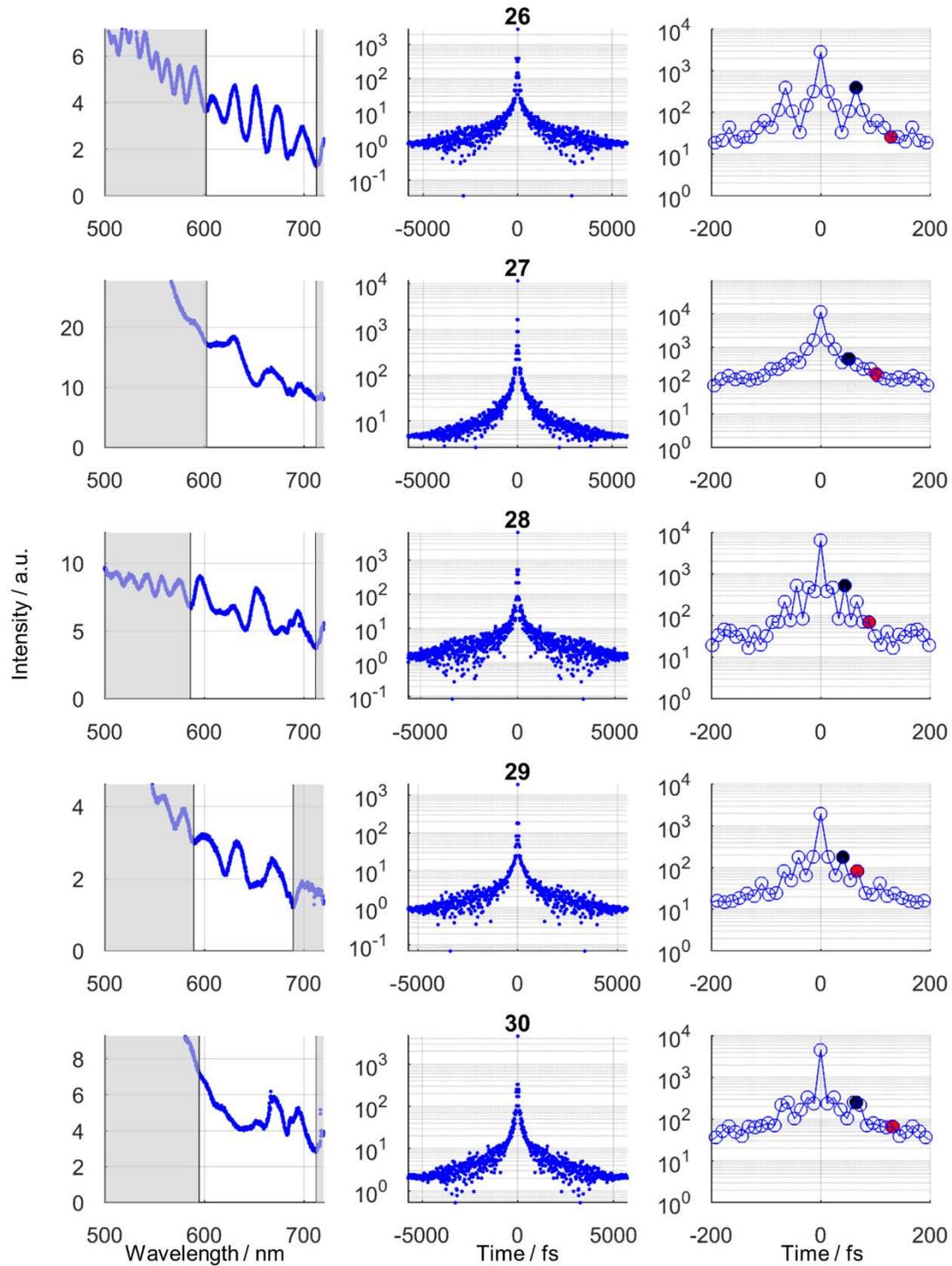

Figure S16. In analogy, the normalized scattering spectrum (left) of silver nanowire with 8 nm silica coating samples 26 to 30, Fourier transformation (middle) and indicated first (right, black dot) and second (right, red dot) plasmon round trip.

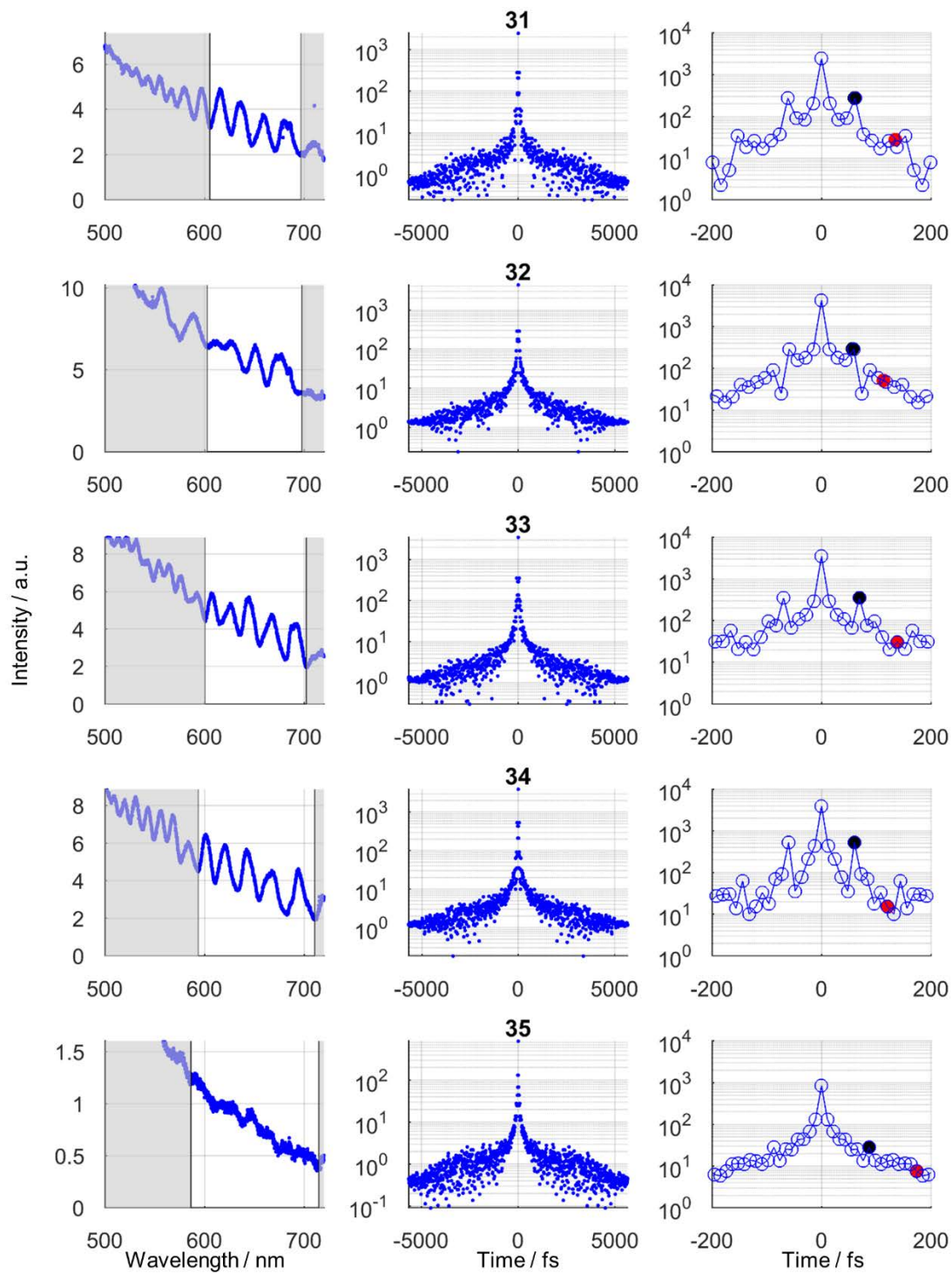

Figure S17. In analogy, the normalized scattering spectrum (left) of silver nanowire with 8 nm silica coating samples 30 to 35, Fourier transformation (middle) and indicated first (right, black dot) and second (right, red dot) plasmon round trip.

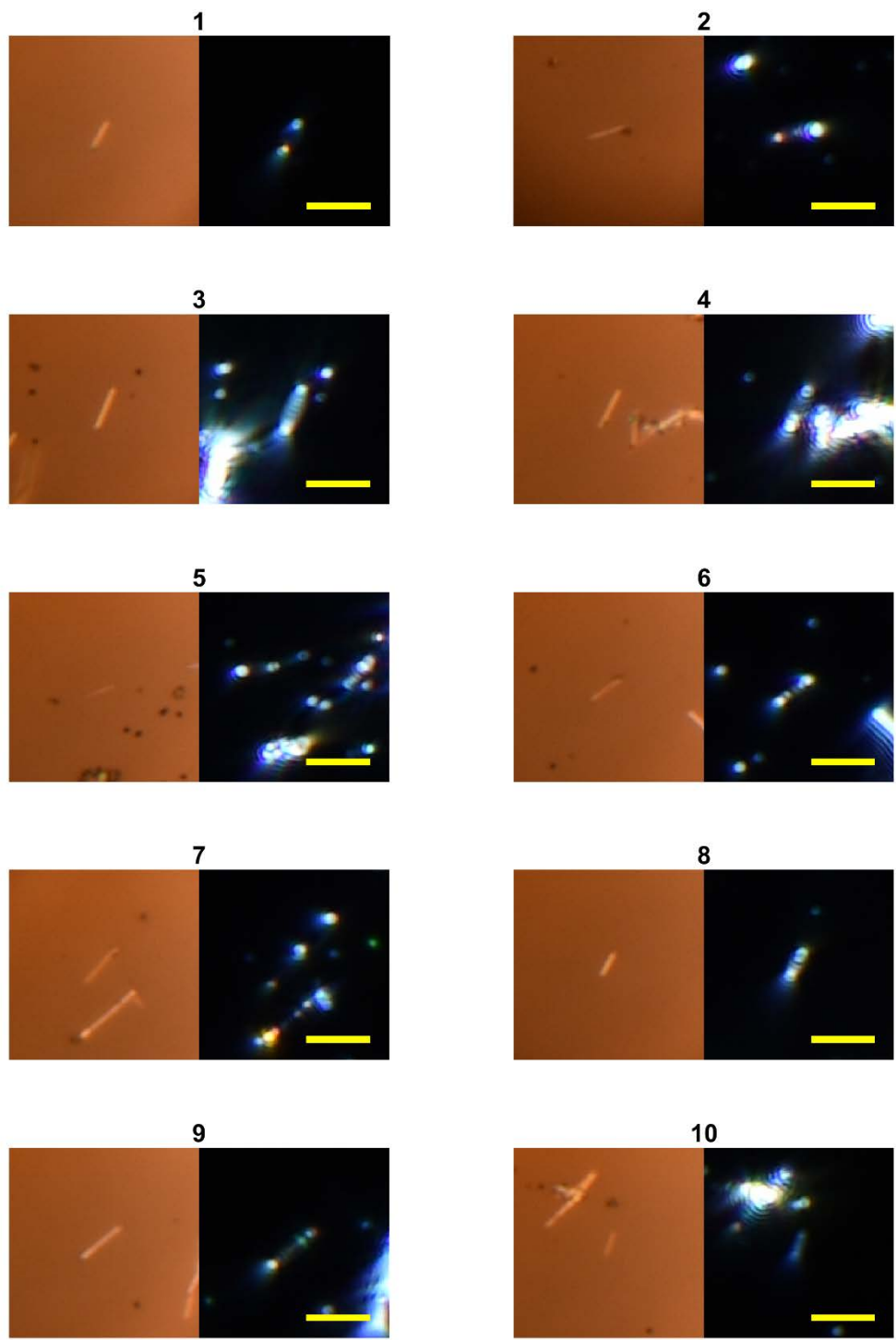

Figure S18. To present limitations in the preparation and the optical method, silver nanowire with 24 nm silica coating samples were studied in bright-field imaging (left) and dark-field imaging (right), respectively. The scale bar is 5 µm.

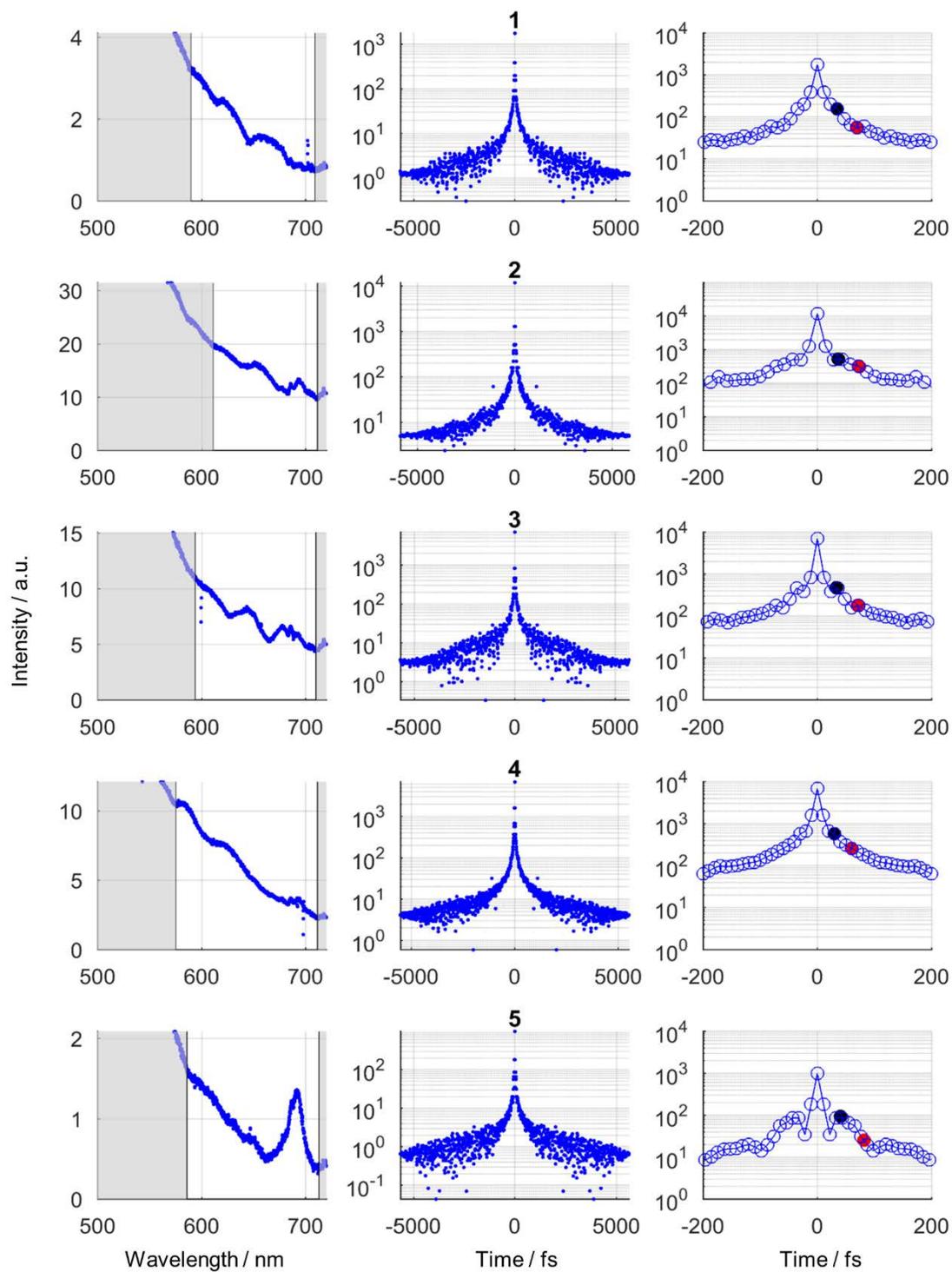

Figure S19. In analogy, the normalized scattering spectrum (left) of silver nanowire with 24 nm silica coating samples 1 to 5, Fourier transformation (middle) and indicated first (right, black dot) and second (right, red dot) plasmon round trip.

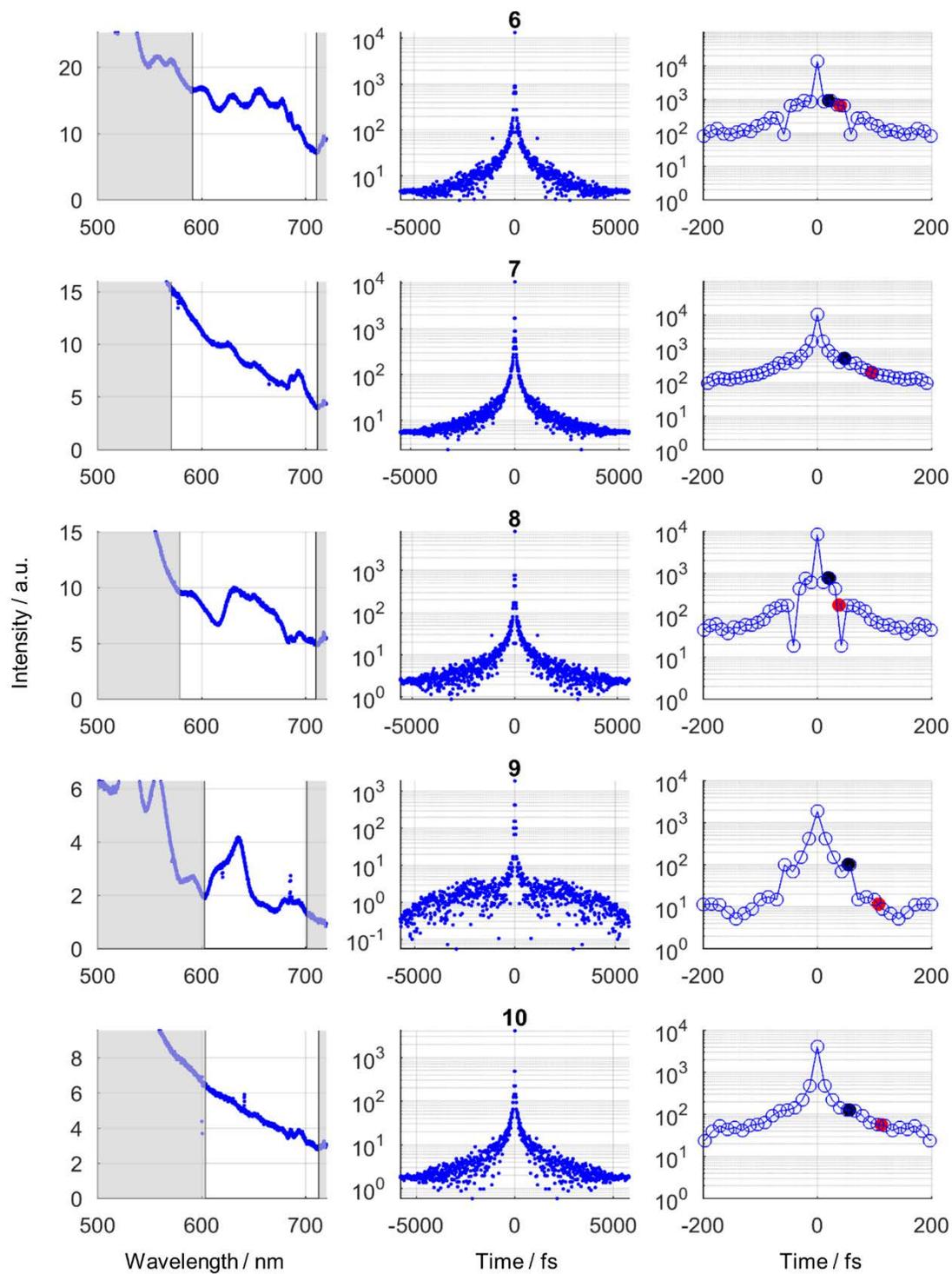

Figure S20. In analogy, the normalized scattering spectrum (left) of silver nanowire with 24 nm silica coating samples 5 to 10, Fourier transformation (middle) and indicated first (right, black dot) and second (right, red dot) plasmon round trip.